\documentclass[aps,prb,reprint,twocolumn,superscriptaddress]{revtex4-1}

\usepackage{graphicx}  
\usepackage{bm}        
\usepackage{amssymb}   
\usepackage{siunitx}    

\usepackage[caption=false]{subfig}  
\usepackage{color}

\usepackage[colorlinks=true, breaklinks=true, bookmarks=false, linkcolor=blue, urlcolor=blue, citecolor=blue,]{hyperref}

\begin{document}

\title{Two-Stage Proximity-Induced Gap-Opening in Topological Insulator--Insulating Ferromagnet (Bi$_x$Sb$_{1-x}$)$_2$Te$_3$--EuS Bilayers}

\author{Qi I. Yang} 
\email{qiyang@cantab.net}
\affiliation{Department of Physics, Stanford University, Stanford, CA 94305, USA}
\affiliation{Geballe Laboratory for Advanced Materials, Stanford University, Stanford, CA 94305, USA}
\affiliation{Stanford Institute for Materials and Energy Sciences, SLAC National Accelerator Laboratory, 2575 Sand Hill Road, Menlo Park, CA 94025, USA}

\author{Aharon Kapitulnik}
\affiliation{Department of Physics, Stanford University, Stanford, CA 94305, USA}
\affiliation{Geballe Laboratory for Advanced Materials, Stanford University, Stanford, CA 94305, USA}
\affiliation{Department of Applied Physics, Stanford University, Stanford, CA 94305, USA}
\date{\today}%
\begin{abstract}%
To further investigate the interplay between ferromagnetism and topological insulators, thin films of the low-carrier topological insulator (Bi$_x$Sb$_{1-x}$)$_2$Te$_3$ were deposited on the insulating ferromagnet EuS (100) \textit{in situ}. AC susceptibility indicates magnetic anomalies between $T\approx30~\mathrm{K}$ and $T\approx60~\mathrm{K}$, well above the Curie temperature $T_C \approx 15~\mathrm{K}$ of EuS. When the Fermi level is close to the Dirac point and the surface state dominates the electric conduction, sharp increases in resistance with decreasing temperatures were observed concurrently with the magnetic anomalies. Positive-negative magnetoresistance crossovers were observed at the Curie temperature, which seem only to appear when the sheet resistance exceeds the Mott-Ioffe-Regel limit $h/e^2$. A two-stage gap-opening process due to magnetic proximity is proposed.
\end{abstract}%
\maketitle%
\tableofcontents
\section{Main Article}\label{sec:main}
Recent studies of topological insulators (TI)~\cite{TI_review, Kane_review}  emphasize their interplay with various forms of magnetism. One of the main objectives is the observation of Quantum Anomalous Hall Effect (QAHE), that is a quantized Hall effect without a magnetic field.~\cite{Haldane1988} Starting from a two-dimensional (2D) TI, also known as a quantum spin Hall system, where a pair of counter-propagating edge states with opposite spins exist, QAHE is realized with the introduction of ferromagnetic order that suppresses one of the spin channels.\cite{ZhangSC_magnetic_impurities,Yu2010} A standard route to achieve a 2D-TI is to reduce the thickness of a 3D-TI until the two opposing surfaces hybridize to form a 2D-TI.  To introduce ferromagnetism, one approach was to dope the bulk 3D-TI with ferromagnetic ions;\cite{Yu2010,XueQK_QAHE, ShenZX_Tl_doping, Cava_doping} while in a second approach, a ferromagnetic layer was brought to contact with the surface of the TI.\cite{bilayer2013, Samarth2013, Moodera2013, Chulkov2017} With sufficiently strong perpendicular anisotropy, time reversal symmetry should be broken at that surface and a QAHE would be realized. To date, a ``true'' QAHE in zero magnetic field was achieved only in magnetically doped TIs,\cite{XueQK_QAHE,Kou2015} and only at temperatures much lower than the ferromagnetic coupling temperature. The need for low temperature has been attributed to the doping-related disorder. While bulk disorder may be alleviated in the bilayer configuration, it is replaced by interface effects.

The first generation of TI--ferromagnet bilayers used bismuth selenide (Bi$_2$Se$_3$) as the TI platform, and EuS~\cite{bilayer2013,Moodera2013} or GdN~\cite{Samarth2013} for the insulating ferromagnet. Relevant to the present study, we previously reported magneto-transport measurements on bilayer samples with europium sulfide (EuS) as the insulating ferromagnet, where a crossover between positive and negative magnetoresistance suggested a proximity effect occurring at the Curie temperature ($T_C$) of EuS.\cite{bilayer2013} Investigating a similar material system, Wei {\it et al.} further detected a low temperature weak hysteresis as a signature for a developing ferromagnetic phase.\cite{Moodera2013} Further investigations by this group, using spin-polarized neutron reflectivity experiments, revealed interfacial magnetism that extended $\sim$2 nm into a $\sim$20 nm Bi$_2$Se$_3$ system, which persisted to temperatures much higher than the $T_C$ of EuS itself.\cite{Moodera2016} While a small increase in $T_C$ of EuS has been reported before, and was attributed to the presence of free bulk carriers,\cite{EuS_ntype, Keller2002} the much larger increase in $T_C$ was attributed solely to an interface effect. However, progress in this bilayer material system has been slow, primarily because of Bi$_2$Se$_3$ quality problems such as interstitials and vacancies, which lift the Fermi level to the bulk conduction band, resulting in n-type bulk conductivity,\cite{Bi2Se3_ARPES1, zhangli2013, Zhanybek3, Fisher2010} thereby complicate the interpretation of experimental results. 

A variety of other 3D-TI materials have been studied in search for an optimal TI platform. In particular, like Bi$_2$Se$_3$, both Bi$_2$Te$_3$ and Sb$_2$Te$_3$ share the same quintuple-layer (QL) crystalline structures with similar lattice constants.\cite{SbStructure, BiStructure} However, unlike Bi$_2$Se$_3$, the Dirac point of either compound is not well exposed in the bulk band gap.\cite{Zhang2009} This was resolved by using the alloy (Bi$_x$Sb$_{1-x}$)$_2$Te$_3$ (BST), introducing a further advantage that electric conduction can be tuned between n-type and p-type by changing the Bi to Sb ratios.\cite{ZhangJS2011} The realization of QAHE by Cr-doping of BST,\cite{Kou2014} exhibiting high sample quality and robust magnetism at low temperatures, which persists even when the film thickness is beyond the 2D hybridization limit, suggests that it should also be tried with a bilayer configuration.

In this paper we present new results on magnetic behavior in the BST--EuS bilayer thin film system. In addition to reproducing similar results as in the Bi$_2$Se$_3$--EuS bilayer system, namely a positive to negative magnetoresistance crossover at the Curie temperature of EuS $T_C\approx15$ K,\cite{bilayer2013} novel magnetic order was observed at the interface between BST and EuS, which persists to $\sim$60 K, much higher than the bulk $T_C$ of EuS. Anomalies in the resistivity and AC Susceptibility suggest a two-stage magnetic proximity induced gap-opening mechanism. In the rest of this paper, the magnetic and transport properties of four representative samples are reported and compared. 

Based on existing procedures,\cite{EuS_PLD, telluride_PLD1, telluride_PLD2, PLD_alt_target2, PLD_alt_target} bilayer samples were fabricated by growing EuS~(100) and BST thin films sequentially \textit{in situ} on Si~(100) substrates by pulsed laser deposition (PLD). Here we present studies on two thin and optimally doped samples (S1 \& S2), where the surface state should dominate the electric conduction;\cite{ZhangJS2011} and, to contrast, two thicker and undoped samples (S3 \& S4), where the Fermi levels intersect the bulk valance band, hence a large contribution of p-type bulk conduction is expected (table.~\ref{tab:samples}). %
\begin{table}[ht]
    \centering
    \begin{ruledtabular}
    \begin{tabular}{l|l|l|l}
        Samples & Ferromagnet & TI Compositions & TI Thicknesses\\
        \hline%
        S1 & EuS (100) & (Bi$_{0.05}$Sb$_{0.95}$)$_2$Te$_3$ & 4~nm\\
        S2 & EuS (100) & (Bi$_{0.05}$Sb$_{0.95}$)$_2$Te$_3$ & 6.5~nm\\
        S3 & EuS (100) & Sb$_2$Te$_3$ & 13~nm\\
        S4 & EuS (100) & Sb$_2$Te$_3$ & 65~nm\\
    \end{tabular}
    \end{ruledtabular}
    \caption{\label{tab:samples}Summary of samples. S1 \& S2 are thin and optimally doped and therefore should have dominant surface conduction; whereas S3 \& S4 are undoped and thicker therefore should have large contribution from the bulk. Composition and thickness are calculated from numbers of laser pulses.}
\end{table}%
X-ray diffraction (XRD) indicates clear (001) orientation of the BST layers. The component of magnetization perpendicular to the films behave similarly to EuS thin films without the TI layer in DC magnetometry.~\footnote{See supplemental material at [URL will be inserted by publisher] for details of sample fabrication and basic characterization, which includes refs.~\onlinecite{EuS_PLD, telluride_PLD1, telluride_PLD2, PLD_alt_target2, PLD_alt_target, ZhangJS2011, SbStructure}.}

While useful as a bulk measurement, DC magnetometry is less suited to detect weak interface phenomena. In particular, measurements above $T_C$ are especially difficult when background interference dominates the SQUID coil centering process.\cite{squid_center_error} AC magnetic susceptibility, on the other hand, has been proven to be very sensitive to thermodynamic transitions as well as surface and local phenomena, as demonstrated in studies of 2D ferromagnetism, spin-glass, superparamagnetism, heavy fermions and superconductivity.\cite{ac_nitroxide, ac_spin_glass, ac_superpara, Ando1994, Gegenwart2005, Schemm2014} To better study the magnetic properties of the interface in a wider temperature range, AC susceptibility of the thin optimally doped sample S1 was measured with a home-made two-coil mutual inductance device\cite{Jeanneret1989,Yazdani1993} at a drive frequency $f=71~\mathrm{kHz}$. The pick-up coil was wound in a gradiometer configuration and mounted inside the drive coil, both casted into a small epoxy cylinder. One end of the cylinder was then polished to allow the sample to be in close proximity to the top of the two concentric coils (see e.g. ref.~\onlinecite{YazdaniThesis}). The same sample was measured in a van der Pauw configuration for DC and Hall resistance measurements. Indeed, where an unusual behavior of the bilayer system is observed, anomalies appear in both susceptibility and resistance measurements.

\begin{figure}[ht]%
    \subfloat{\label{fig:s1rvt}}%
    \subfloat{\label{fig:s1mi}}%
    \includegraphics[width=\columnwidth]{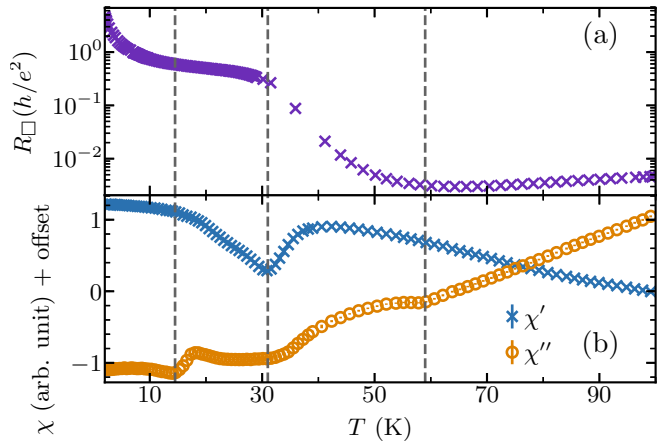}%
    \caption{\label{mirvt}Temperature dependence of (a)~sheet resistance, and (b)~AC magnetic susceptibility of sample S1 in zero magnetic field. When an unusual behavior of the bilayer system is observed, anomalies appear in both susceptibility and resistance.}%
\end{figure}

\begin{figure}[ht]%
    \centering%
    \subfloat{\label{fig:mi0t0}}%
    \subfloat{\label{fig:mi0t40}}%
    \subfloat{\label{fig:mi0.02t0}}%
    \subfloat{\label{fig:mi0.02t40}}%
    \subfloat{\label{fig:mi2t0}}%
    \subfloat{\label{fig:mi2t40}}%
    \includegraphics[width=\linewidth]{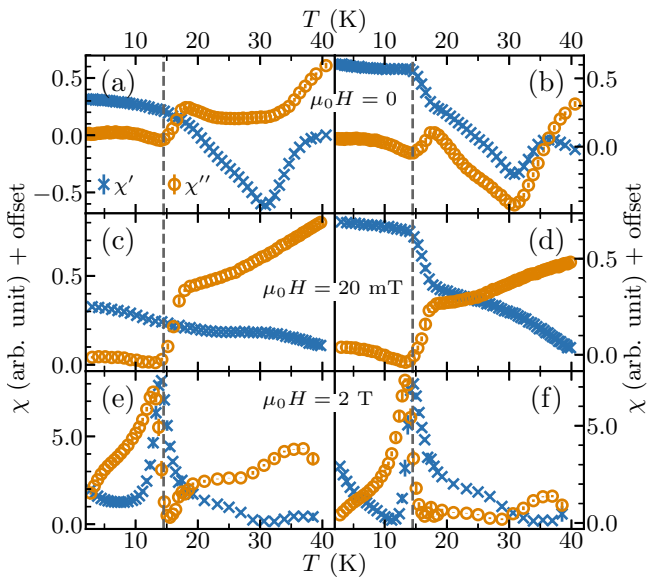}%
    \caption{\label{fig:mi}(Color online)~Real ($\chi'$, crosses) and imaginary ($\chi''$, circles) parts of the AC susceptibility of sample S1 as functions of temperatures close to the Curie temperature (dashed lines). (a, b)~In zero field, (c, d)~in 20~mT and (e, f)~in 2~T DC fields perpendicular to the film. The left column shows as-measured data whereas the right column includes $40^{\circ}$ phase rotations. Error bars in all figures in this paper represent the estimated 95~\protect\% confidence intervals.}%
\end{figure}

A striking example for the correspondence between the zero field AC susceptibility and DC resistance is shown in fig.~\ref{mirvt}. %
This 4~nm sample is expected to be very close to the 2D-TI r\'egime where magnetism from the proximitized EuS is expected to have maximum effect. There is a clear effect at the Curie temperature of EuS ($\sim$15 K), where the sheet resistance starts its low-temperature increase, while the the AC susceptibility saturates in magnitude. However, these expected effects are just the last of the magnetic response as we lower the temperature. A dramatic increase in resistance, associated with a cusp in the imaginary part of AC susceptibility, is first observed at 60 K. Lowering the temperature, the sheet resistance seems to almost saturate at $\sim$30 K, at which point the real (inductive) part of the susceptibility shows a dip and the imaginary (dissipative) part almost saturates. Such anomaly seems to be readily suppressed by a small perpendicular magnetic field (fig.~\ref{fig:mi0.02t0}), which is consistent with a change in the magnetic configuration at the interface, such as that proposed in ref.~\onlinecite{Moodera2016}. %
In a strong perpendicular DC magnetic field, where the magnetization in the ferromagnetic phase is forced to align with the applied field similarly to ferromagnets measured on their easy axes, the real and imaginary parts of the AC susceptibility should exhibit peaks just above and below $T_C$ respectively.\cite{ac_nitroxide, Venus2004} However the as-measured data slightly deviate from such expected behavior (fig.~\ref{fig:mi2t0}). This is likely due to the phase rotation and complex offset introduced by the finite resistance, capacitance and inductance in the wiring of the cryostat and instruments. Indeed the expected behavior is recovered by applying a $40^{\circ}$ phase rotation (fig.~\ref{fig:mi2t40}). For comparison, the AC susceptibility in zero and 20~mT DC fields are also presented with $40^{\circ}$ phase rotations in the right column in fig.~\ref{fig:mi} next to their as-measured counterparts. In particular, in 20~mT DC field, where the magnetization is mostly in-plane and the anomaly above $T_C$ is suppressed, the AC susceptibility after phase rotation (fig.~\ref{fig:mi0.02t40}) also roughly conforms with the expected behavior of a thin film ferromagnet measured on its hard axis.\cite{Jensen2003}

\begin{figure}[ht]%
    \centering%
    \includegraphics[width=\linewidth]{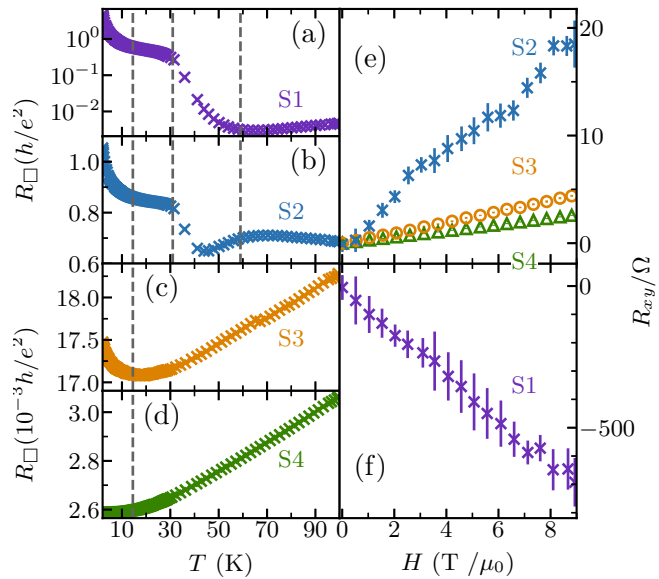}%
    \subfloat{\label{fig:rvt_s1}}%
    \subfloat{\label{fig:rvt_s2}}%
    \subfloat{\label{fig:rvt_s3}}%
    \subfloat{\label{fig:rvt_s4}}%
    \subfloat{\label{fig:hallp}}%
    \subfloat{\label{fig:halln}}%
    \caption{\label{fig:rvt}(Color online)~Resistive anomalies observed in samples (\protect\subref*{fig:rvt_s1})~S1 and (\protect\subref*{fig:rvt_s2})~S2 at the same temperatures where magnetic anomalies occur, but not in those with intrinsic thicker TI layers (\protect\subref*{fig:rvt_s3})~S3 and (\protect\subref*{fig:rvt_s4})~S4. (\protect\subref*{fig:hallp},~\protect\subref*{fig:halln}) The Hall effect indicates decreasing carrier densities per unit area from S4 to S1 and a shift from p-type to n-type.}%
\end{figure}

Transport data are shown in fig.~\ref{fig:rvt}, summarizing the zero-field sheet resistance (figs.~\ref{fig:rvt_s1}--\ref{fig:rvt_s4}) and the Hall resistance (figs.~\ref{fig:hallp} \& \ref{fig:halln}) of the four bilayer samples. %
The Hall effect indicates that S2--S4 have holes as the majority carrier, whereas S1 exhibits electron character. A possible reason for a change in majority carrier types between S2 and S1, from holes to electrons, could be the reduction in thickness, hence a stronger influence from the chemical potential of the EuS layer, which has a natural tendency to have electron donors.\cite{EuS_ntype} While a small elevation of chemical potential may not produce measurable electric conduction in EuS due to its large band gap,\cite{EuS_band1, EuS_band2} in the BST layer, however, if the Fermi level is below and very close to the Dirac point,\cite{ZhangJS2011} where excitations exhibit electron-hole symmetry, even a small elevation may change the majority carrier type. Similarly to S1, a resistive transition was observed in the slightly thicker optimally doped sample S2 (fig.~\ref{fig:rvt_s2}) near $T\approx30~\mathrm{K}$. Such resistive transitions were neither observed in samples S3 \& S4 (figs.~\ref{fig:rvt_s3} \& \ref{fig:rvt_s4}) nor in the Bi$_2$Se$_3$--EuS bilayers in ref.~\onlinecite{bilayer2013}, where in both cases the ferromagnetism is present but the bulk conduction is more dominant; nor in BST samples near the optimal doping level reported in ref.~\onlinecite{ZhangJS2011}, where the surface conduction dominates but in absence of magnetism. These strongly suggest that the resistive transition observed is a result of proximity between the magnetic order and the surface state. Indeed, the interface magnetization is expected to open a gap at the TI's surface state, hence reduce its contribution to the overall conduction, which would only be observed when the EuS layer is highly insulating and the surface state dominates the conduction in the TI layer.

\begin{figure}[ht]%
    \centering%
    \includegraphics[width=\linewidth]{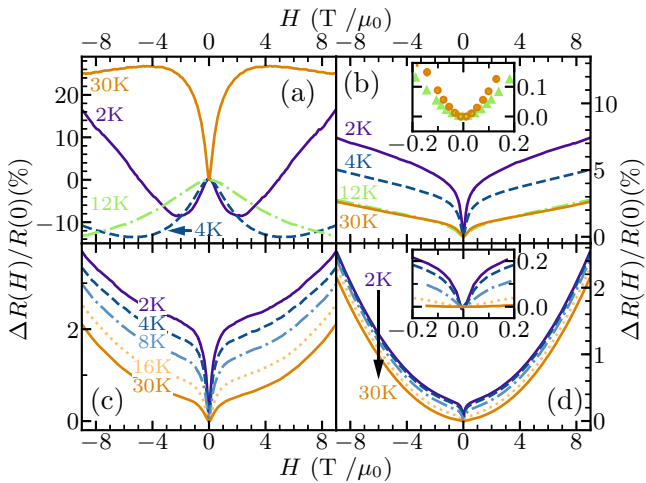}%
    \subfloat{\label{fig:mr_s1}}%
    \subfloat{\label{fig:mr_s2}}%
    \subfloat{\label{fig:mr_s3}}%
    \subfloat{\label{fig:mr_s4}}%
    \caption{\label{fig:mr}(Color online)~Magnetoresistance at representative temperatures of (\protect\subref*{fig:mr_s1})~S1, (\protect\subref*{fig:mr_s2})~S2, (\protect\subref*{fig:mr_s3})~S3 and (\protect\subref*{fig:mr_s4})~S4. (\protect\subref*{fig:mr_s2}: insert) Low-field behavior of S2 at $T = 12\mathrm{K}$ (triangles) and at $T = 30\mathrm{K}$ (circles), showing reverse temperature dependence. (\protect\subref*{fig:mr_s4}: insert) Low-field features of S4.}%
\end{figure}

\begin{figure}[ht]%
    \centering
    \includegraphics[width=\linewidth]{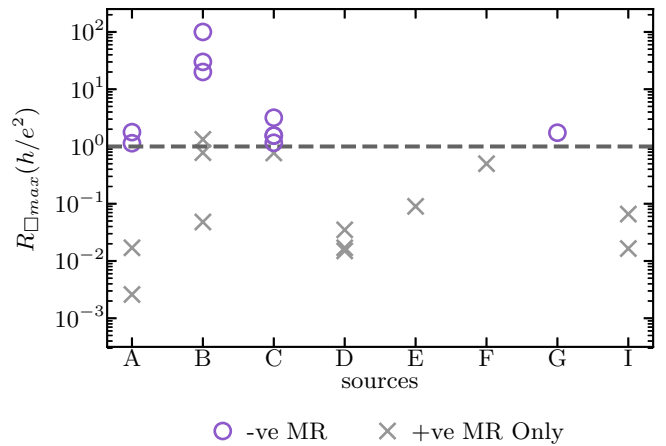}
    \caption{\label{fig:wl_trend} The maximum sheet resistance of bilayer samples at zero magnetic field from a variety of sources (A: this paper, B: ref.~\protect\onlinecite{bilayer2013} and unpublished data, C--I: refs.~\protect\onlinecite{Samarth2017, Shi2014, Petta2014, Wang2014, Tian2016, Qiu2017}) The Mott-Ioffe-Regel limit ($R_\Box = h/e^2$) seems to separate samples showing signatures of negative MR below $T_C$ (violet circles), and those only display positive MR (gray crosses)}
\end{figure}

The magnetoresistance (MR) of the bilayer samples was measured at representative temperatures and presented in fig.~\ref{fig:mr}. %
A positive to negative MR crossover at $T_C$ was observed in S1 (fig.~\ref{fig:mr_s1}), similarly to previously reported behavior of thin Bi$_2$Se$_3$--EuS bilayers.\cite{bilayer2013} Above $T_C$, a sharp positive MR feature exists near zero field as ubiquitously observed in TI thin films; whereas below $T_C$ a negative MR emerges. In S2 the MR remains positive at all measured temperatures (fig.~\ref{fig:mr_s2}), however the low-field feature is broader at 12~K than at 30~K (fig.~\ref{fig:mr_s2}: insert), suggesting a developing negative component, similar to Bi$_2$Se$_3$--EuS bilayers close to $T_C$.\cite{bilayer2013} In thicker undoped samples S3~\&~S4 (figs.~\ref{fig:mr_s3}~\&~\ref{fig:mr_s4}), only positive MR was observed, which sharpens monotonously with decreasing temperature, in addition to parabolic backgrounds that are typically observed in thicker TI films.\cite{TI_WAL_thickness} While in our previous studies of Bi$_2$Se$_3$-EuS bilayers the Fermi levels were likely well inside the bulk conduction band, and therefore the mechanism of the emergent negative MR remained inconclusive; in the present study, specifically for samples S1 and S2, the doping levels and the Hall effects suggest that the Fermi levels are very close to the Dirac point and well inside the bulk band gap. This case was studied theoretically, suggesting that either gap-opening at the Dirac point~\cite{Glazman2012, Lu2011} or coexistence of ferromagnetism and spin-orbit coupling~\cite{WL_ferromagnetism} as the origin for the negative MR. Finally, we note that the crossovers from positive to negative MR have also been observed in bilayer structures with different TIs and ferromagnets,\cite{Samarth2017, Tian2016} interestingly only when the sheet resistance exceeds the Mott-Ioffe-Regel limit~\cite{Mott_book, Fradkin1986b} in two-dimensions $h/e^2$ (fig.~\ref{fig:wl_trend}). %
While most available theories on magneto-transport properties of TI thin films have been formulated in terms of weak localization, we note that, being an orbital quantum interference effect, the concept of weak localization is not easily applicable in such r\'egime. 

To summarize, (Bi$_x$Sb$_{1-x}$)$_2$Te$_3$--EuS bilayers were fabricated by pulsed laser deposition. AC magnetic susceptibility displayed anomalies well above the bulk $T_C$ of EuS. Resistive transitions were observed concurrently with magnetic anomalies in thin optimally doped samples, where the Fermi levels are close to the Dirac point, suggesting a gap opened at the Dirac point at the interface. Similarly to previous results, negative magnetoresistance was observed below $T_C$ near zero fields whereas positive magnetoresistance was recovered above $T_C$. Together these suggest a two-stage gap-opening mechanism at the TI surface state Dirac point as result of proximity to an insulating ferromagnet.
\begin{acknowledgments}
We thank Jiecheng Zhang, Sejoon Lim and Shuai Shao for helpful discussions. The mutual inductance coil was crafted by Alan Fang. Sample fabrication and characterization were partly performed at the Stanford Nano Shared Facilities (SNSF), supported by the National Science Foundation under award ECCS-1542152.  This work is supported by the Department of Energy,  Office of Science, Basic Energy Sciences, Materials Sciences and Engineering Division, under Contract DE-AC02-76SF00515. Initial work was supported by DARPA, MesoDynamic Architecture Program (MESO) through the contract number N66001-11-1-4105.
\end{acknowledgments}
\section{Supplemental Material: Sample Fabrication and Characterization}\label{sec:SM}
Samples of (Bi$_{x}$Sb$_{1-x}$)$_2$Te$_3$--EuS bilayer thin films were fabricated by pulsed laser deposition (PLD). For the ferromagnet layer, 40~nm EuS thin films were grown on intrinsic Si (100) substrates with native oxide. With the previously reported recipe,\cite{EuS_PLD} high quality EuS thin films were consistently obtained, characterized by single (100) orientations, atomically smooth surfaces, immeasurably high sheet resistance, and magnetic anisotropy exhibiting an out-of plane component of the magnetizations. To optimize the quality of the interfaces, the TI layer was subsequently grown by PLD \textit{in situ}. Based on existing reports on both Sb$_2$Te$_3$ and Bi$_2$Te$_3$,\cite{telluride_PLD1, telluride_PLD2} we established a procedure to deposit (Bi$_{x}$Sb$_{1-x}$)$_2$Te$_3$ by alternating the targets of the two compounds.\cite{PLD_alt_target2, PLD_alt_target} Following each EuS deposition carried out in high vacuum of $\sim10^{-7}~\mathrm{torr}$,  the sample was allowed to cool to $\SI{300}{\degreeCelsius}$ before 200~millitorr of argon gas mixed with 2\% hydrogen was introduced to diffuse the plasma plumes and to prevent potential oxidation from residual gases. Sputtering targets (Kurt J. Lesker, 99.999\%) were ablated $\SI{5}{\cm}$ away from the sample with 25~ns 248~nm KrF excimer laser pulses at 0.55 $\mathrm{J\cdot{}cm^{-2}}$ fluence and 5~Hz repetition rate. The thickness of a $\sim\SI{40}{nm}$ film was measured by atomic force profiliometry and the average deposition rate was $\SI{0.22}{\angstrom}$/pulse. To achieve the optimal composition (5$\%$ bismuth doping), where the Fermi level is inside the bulk band gap and closest to the Dirac point,\cite{ZhangJS2011} the Bi$_2$Te$_3$ target was ablated by one pulse once per 19 pulses on Sb$_2$Te$_3$.

\begin{figure}[ht]%
    \centering%
    \includegraphics[width=\linewidth]{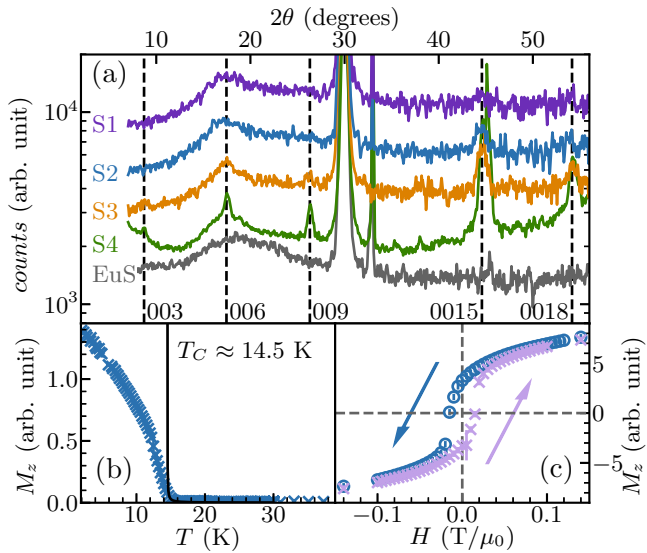}%
    \subfloat{\label{fig:xrd}}%
    \subfloat{\label{fig:mvt}}%
    \subfloat{\label{fig:mvh}}%
    \caption{\label{fig:material}(Color online)~(\protect\subref*{fig:xrd})~Semi-log X-ray diffraction spectra of the four (Bi$_{x}$Sb$_{1-x}$)$_2$Te$_3$--EuS bilayer samples, with thickness of the TI layer increasing from the top to the bottom, compared to a EuS-only thin film of similar thickness. K-$\beta$ spectral contamination exists in the spectrum of sample S4 due to unavailable monochromator. Dashed lines mark the expected positions of the (Bi,Sb)$_2$Te$_3$ [001] peaks.\protect\cite{SbStructure} Magnetization of sample S2 as functions of (\protect\subref*{fig:mvt})~temperatures and (\protect\subref*{fig:mvh})~perpendicular magnetic fields (arrows indication field sweep directions). A fitting to the Curie-Weiss law is shown as the black solid curve. Error bars represent the estimated 95~\protect\% confidence intervals.}%
\end{figure}

The four samples studied in the main paper are described in table.~\ref{tab:samples}. %
\begin{table}[ht]
    \centering
    \begin{ruledtabular}
    \begin{tabular}{l|l|l|l}
        Samples & Ferromagnet & TI Compositions & TI Thicknesses\\
        \hline
        S1 & EuS (100) & (Bi$_{0.05}$Sb$_{0.95}$)$_2$Te$_3$ & 4~nm\\
        S2 & EuS (100) & (Bi$_{0.05}$Sb$_{0.95}$)$_2$Te$_3$ & 6.5~nm\\
        S3 & EuS (100) & Sb$_2$Te$_3$ & 13~nm\\
        S4 & EuS (100) & Sb$_2$Te$_3$ & 65~nm\\
    \end{tabular}
    \end{ruledtabular}
    \caption{\label{tab:samples}Summary of samples. Composition and thickness are calculated from numbers of laser pulses.}
\end{table}
The X-ray diffraction (XRD) spectra of these samples (fig.~\ref{fig:xrd}) indicate clear (001) orientations of the (Bi$_{x}$Sb$_{1-x}$)$_2$Te$_3$ layers. %
Their bulk magnetic properties were studied with a superconducting quantum interference device (SQUID) magnetometer. Examples are presented in figs.~\ref{fig:mvt}~\&~\ref{fig:mvh} for sample S2. The sample was cooled in zero magnetic field to $T=2~\mathrm{K}$, at which the centering procedure of SQUID was carried out. Subsequently, the field-dependence of the magnetization was measured in an external magnetic field perpendicular to the thin film sweeping between $\mu_0H = +2~\mathrm{T}$ and $\mu_0H = -2~\mathrm{T}$ (fig.~\ref{fig:mvh}). Finally the external field was reduced to zero from $\mu_0H = +2~\mathrm{T}$, and the temperature dependence of the magnetization was measured during warming up in zero field (fig.~\ref{fig:mvt}). The component of magnetization perpendicular to the films behaves similarly to EuS thin films without the TI layer.\cite{EuS_PLD} By fitting to the Curie-Weiss law in the paramagnetic r\'egime, the Curie temperature was determined to be $T_C = 14.5\pm0.3\mathrm{K}$.

\bibliography{bilayer2018}

\begin{thebibliography}{57}%
\makeatletter
\providecommand \@ifxundefined [1]{%
 \@ifx{#1\undefined}
}%
\providecommand \@ifnum [1]{%
 \ifnum #1\expandafter \@firstoftwo
 \else \expandafter \@secondoftwo
 \fi
}%
\providecommand \@ifx [1]{%
 \ifx #1\expandafter \@firstoftwo
 \else \expandafter \@secondoftwo
 \fi
}%
\providecommand \natexlab [1]{#1}%
\providecommand \enquote  [1]{``#1''}%
\providecommand \bibnamefont  [1]{#1}%
\providecommand \bibfnamefont [1]{#1}%
\providecommand \citenamefont [1]{#1}%
\providecommand \href@noop [0]{\@secondoftwo}%
\providecommand \href [0]{\begingroup \@sanitize@url \@href}%
\providecommand \@href[1]{\@@startlink{#1}\@@href}%
\providecommand \@@href[1]{\endgroup#1\@@endlink}%
\providecommand \@sanitize@url [0]{\catcode `\\12\catcode `\$12\catcode
  `\&12\catcode `\#12\catcode `\^12\catcode `\_12\catcode `\%12\relax}%
\providecommand \@@startlink[1]{}%
\providecommand \@@endlink[0]{}%
\providecommand \url  [0]{\begingroup\@sanitize@url \@url }%
\providecommand \@url [1]{\endgroup\@href {#1}{\urlprefix }}%
\providecommand \urlprefix  [0]{URL }%
\providecommand \Eprint [0]{\href }%
\providecommand \doibase [0]{http://dx.doi.org/}%
\providecommand \selectlanguage [0]{\@gobble}%
\providecommand \bibinfo  [0]{\@secondoftwo}%
\providecommand \bibfield  [0]{\@secondoftwo}%
\providecommand \translation [1]{[#1]}%
\providecommand \BibitemOpen [0]{}%
\providecommand \bibitemStop [0]{}%
\providecommand \bibitemNoStop [0]{.\EOS\space}%
\providecommand \EOS [0]{\spacefactor3000\relax}%
\providecommand \BibitemShut  [1]{\csname bibitem#1\endcsname}%
\let\auto@bib@innerbib\@empty
\bibitem [{\citenamefont {Qi}\ and\ \citenamefont {Zhang}(2011)}]{TI_review}%
  \BibitemOpen
  \bibfield  {author} {\bibinfo {author} {\bibfnamefont {X.-L.}\ \bibnamefont
  {Qi}}\ and\ \bibinfo {author} {\bibfnamefont {S.-C.}\ \bibnamefont {Zhang}},\
  }\href {\doibase 10.1103/RevModPhys.83.1057} {\bibfield  {journal} {\bibinfo
  {journal} {Rev. Mod. Phys.}\ }\textbf {\bibinfo {volume} {83}},\ \bibinfo
  {pages} {1057} (\bibinfo {year} {2011})}\BibitemShut {NoStop}%
\bibitem [{\citenamefont {Hasan}\ and\ \citenamefont
  {Kane}(2010)}]{Kane_review}%
  \BibitemOpen
  \bibfield  {author} {\bibinfo {author} {\bibfnamefont {M.~Z.}\ \bibnamefont
  {Hasan}}\ and\ \bibinfo {author} {\bibfnamefont {C.~L.}\ \bibnamefont
  {Kane}},\ }\href {\doibase 10.1103/RevModPhys.82.3045} {\bibfield  {journal}
  {\bibinfo  {journal} {Rev. Mod. Phys.}\ }\textbf {\bibinfo {volume} {82}},\
  \bibinfo {pages} {3045} (\bibinfo {year} {2010})}\BibitemShut {NoStop}%
\bibitem [{\citenamefont {Haldane}(1988)}]{Haldane1988}%
  \BibitemOpen
  \bibfield  {author} {\bibinfo {author} {\bibfnamefont {F.~D.~M.}\
  \bibnamefont {Haldane}},\ }\href {\doibase 10.1103/PhysRevLett.61.2015}
  {\bibfield  {journal} {\bibinfo  {journal} {Phys. Rev. Lett.}\ }\textbf
  {\bibinfo {volume} {61}},\ \bibinfo {pages} {2015} (\bibinfo {year}
  {1988})}\BibitemShut {NoStop}%
\bibitem [{\citenamefont {Liu}\ \emph {et~al.}(2009)\citenamefont {Liu},
  \citenamefont {Liu}, \citenamefont {Xu}, \citenamefont {Qi},\ and\
  \citenamefont {Zhang}}]{ZhangSC_magnetic_impurities}%
  \BibitemOpen
  \bibfield  {author} {\bibinfo {author} {\bibfnamefont {Q.}~\bibnamefont
  {Liu}}, \bibinfo {author} {\bibfnamefont {C.-X.}\ \bibnamefont {Liu}},
  \bibinfo {author} {\bibfnamefont {C.}~\bibnamefont {Xu}}, \bibinfo {author}
  {\bibfnamefont {X.-L.}\ \bibnamefont {Qi}}, \ and\ \bibinfo {author}
  {\bibfnamefont {S.-C.}\ \bibnamefont {Zhang}},\ }\href {\doibase
  10.1103/PhysRevLett.102.156603} {\bibfield  {journal} {\bibinfo  {journal}
  {Phys. Rev. Lett.}\ }\textbf {\bibinfo {volume} {102}},\ \bibinfo {pages}
  {156603} (\bibinfo {year} {2009})}\BibitemShut {NoStop}%
\bibitem [{\citenamefont {Yu}\ \emph {et~al.}(2010)\citenamefont {Yu},
  \citenamefont {Zhang}, \citenamefont {Zhang}, \citenamefont {Zhang},
  \citenamefont {Dai},\ and\ \citenamefont {Fang}}]{Yu2010}%
  \BibitemOpen
  \bibfield  {author} {\bibinfo {author} {\bibfnamefont {R.}~\bibnamefont
  {Yu}}, \bibinfo {author} {\bibfnamefont {W.}~\bibnamefont {Zhang}}, \bibinfo
  {author} {\bibfnamefont {H.-J.}\ \bibnamefont {Zhang}}, \bibinfo {author}
  {\bibfnamefont {S.-C.}\ \bibnamefont {Zhang}}, \bibinfo {author}
  {\bibfnamefont {X.}~\bibnamefont {Dai}}, \ and\ \bibinfo {author}
  {\bibfnamefont {Z.}~\bibnamefont {Fang}},\ }\href {\doibase
  10.1126/science.1187485} {\bibfield  {journal} {\bibinfo  {journal}
  {Science}\ }\textbf {\bibinfo {volume} {329}},\ \bibinfo {pages} {61}
  (\bibinfo {year} {2010})}\BibitemShut {NoStop}%
\bibitem [{\citenamefont {Chang}\ \emph {et~al.}(2013)\citenamefont {Chang},
  \citenamefont {Zhang}, \citenamefont {Feng}, \citenamefont {Shen},
  \citenamefont {Zhang}, \citenamefont {Guo}, \citenamefont {Li}, \citenamefont
  {Ou}, \citenamefont {Wei}, \citenamefont {Wang}, \citenamefont {Ji},
  \citenamefont {Feng}, \citenamefont {Ji}, \citenamefont {Chen}, \citenamefont
  {Jia}, \citenamefont {Dai}, \citenamefont {Fang}, \citenamefont {Zhang},
  \citenamefont {He}, \citenamefont {Wang}, \citenamefont {Lu}, \citenamefont
  {Ma},\ and\ \citenamefont {Xue}}]{XueQK_QAHE}%
  \BibitemOpen
  \bibfield  {author} {\bibinfo {author} {\bibfnamefont {C.-Z.}\ \bibnamefont
  {Chang}}, \bibinfo {author} {\bibfnamefont {J.}~\bibnamefont {Zhang}},
  \bibinfo {author} {\bibfnamefont {X.}~\bibnamefont {Feng}}, \bibinfo {author}
  {\bibfnamefont {J.}~\bibnamefont {Shen}}, \bibinfo {author} {\bibfnamefont
  {Z.}~\bibnamefont {Zhang}}, \bibinfo {author} {\bibfnamefont
  {M.}~\bibnamefont {Guo}}, \bibinfo {author} {\bibfnamefont {K.}~\bibnamefont
  {Li}}, \bibinfo {author} {\bibfnamefont {Y.}~\bibnamefont {Ou}}, \bibinfo
  {author} {\bibfnamefont {P.}~\bibnamefont {Wei}}, \bibinfo {author}
  {\bibfnamefont {L.-L.}\ \bibnamefont {Wang}}, \bibinfo {author}
  {\bibfnamefont {Z.-Q.}\ \bibnamefont {Ji}}, \bibinfo {author} {\bibfnamefont
  {Y.}~\bibnamefont {Feng}}, \bibinfo {author} {\bibfnamefont {S.}~\bibnamefont
  {Ji}}, \bibinfo {author} {\bibfnamefont {X.}~\bibnamefont {Chen}}, \bibinfo
  {author} {\bibfnamefont {J.}~\bibnamefont {Jia}}, \bibinfo {author}
  {\bibfnamefont {X.}~\bibnamefont {Dai}}, \bibinfo {author} {\bibfnamefont
  {Z.}~\bibnamefont {Fang}}, \bibinfo {author} {\bibfnamefont {S.-C.}\
  \bibnamefont {Zhang}}, \bibinfo {author} {\bibfnamefont {K.}~\bibnamefont
  {He}}, \bibinfo {author} {\bibfnamefont {Y.}~\bibnamefont {Wang}}, \bibinfo
  {author} {\bibfnamefont {L.}~\bibnamefont {Lu}}, \bibinfo {author}
  {\bibfnamefont {X.-C.}\ \bibnamefont {Ma}}, \ and\ \bibinfo {author}
  {\bibfnamefont {Q.-K.}\ \bibnamefont {Xue}},\ }\href {\doibase
  10.1126/science.1234414} {\bibfield  {journal} {\bibinfo  {journal}
  {Science}\ }\textbf {\bibinfo {volume} {340}},\ \bibinfo {pages} {167}
  (\bibinfo {year} {2013})}\BibitemShut {NoStop}%
\bibitem [{\citenamefont {Chen}\ \emph {et~al.}(2010)\citenamefont {Chen},
  \citenamefont {Chu}, \citenamefont {Analytis}, \citenamefont {Liu},
  \citenamefont {Igarashi}, \citenamefont {Kuo}, \citenamefont {Qi},
  \citenamefont {Mo}, \citenamefont {Moore}, \citenamefont {Lu}, \citenamefont
  {Hashimoto}, \citenamefont {Sasagawa}, \citenamefont {Zhang}, \citenamefont
  {Fisher}, \citenamefont {Hussain},\ and\ \citenamefont
  {Shen}}]{ShenZX_Tl_doping}%
  \BibitemOpen
  \bibfield  {author} {\bibinfo {author} {\bibfnamefont {Y.~L.}\ \bibnamefont
  {Chen}}, \bibinfo {author} {\bibfnamefont {J.-H.}\ \bibnamefont {Chu}},
  \bibinfo {author} {\bibfnamefont {J.~G.}\ \bibnamefont {Analytis}}, \bibinfo
  {author} {\bibfnamefont {Z.~K.}\ \bibnamefont {Liu}}, \bibinfo {author}
  {\bibfnamefont {K.}~\bibnamefont {Igarashi}}, \bibinfo {author}
  {\bibfnamefont {H.-H.}\ \bibnamefont {Kuo}}, \bibinfo {author} {\bibfnamefont
  {X.~L.}\ \bibnamefont {Qi}}, \bibinfo {author} {\bibfnamefont {S.~K.}\
  \bibnamefont {Mo}}, \bibinfo {author} {\bibfnamefont {R.~G.}\ \bibnamefont
  {Moore}}, \bibinfo {author} {\bibfnamefont {D.~H.}\ \bibnamefont {Lu}},
  \bibinfo {author} {\bibfnamefont {M.}~\bibnamefont {Hashimoto}}, \bibinfo
  {author} {\bibfnamefont {T.}~\bibnamefont {Sasagawa}}, \bibinfo {author}
  {\bibfnamefont {S.~C.}\ \bibnamefont {Zhang}}, \bibinfo {author}
  {\bibfnamefont {I.~R.}\ \bibnamefont {Fisher}}, \bibinfo {author}
  {\bibfnamefont {Z.}~\bibnamefont {Hussain}}, \ and\ \bibinfo {author}
  {\bibfnamefont {Z.~X.}\ \bibnamefont {Shen}},\ }\href {\doibase
  10.1126/science.1189924} {\bibfield  {journal} {\bibinfo  {journal}
  {Science}\ }\textbf {\bibinfo {volume} {329}},\ \bibinfo {pages} {659}
  (\bibinfo {year} {2010})}\BibitemShut {NoStop}%
\bibitem [{\citenamefont {Hor}\ \emph {et~al.}(2010)\citenamefont {Hor},
  \citenamefont {Roushan}, \citenamefont {Beidenkopf}, \citenamefont {Seo},
  \citenamefont {Qu}, \citenamefont {Checkelsky}, \citenamefont {Wray},
  \citenamefont {Hsieh}, \citenamefont {Xia}, \citenamefont {Xu}, \citenamefont
  {Qian}, \citenamefont {Hasan}, \citenamefont {Ong}, \citenamefont {Yazdani},\
  and\ \citenamefont {Cava}}]{Cava_doping}%
  \BibitemOpen
  \bibfield  {author} {\bibinfo {author} {\bibfnamefont {Y.~S.}\ \bibnamefont
  {Hor}}, \bibinfo {author} {\bibfnamefont {P.}~\bibnamefont {Roushan}},
  \bibinfo {author} {\bibfnamefont {H.}~\bibnamefont {Beidenkopf}}, \bibinfo
  {author} {\bibfnamefont {J.}~\bibnamefont {Seo}}, \bibinfo {author}
  {\bibfnamefont {D.}~\bibnamefont {Qu}}, \bibinfo {author} {\bibfnamefont
  {J.~G.}\ \bibnamefont {Checkelsky}}, \bibinfo {author} {\bibfnamefont
  {L.~A.}\ \bibnamefont {Wray}}, \bibinfo {author} {\bibfnamefont
  {D.}~\bibnamefont {Hsieh}}, \bibinfo {author} {\bibfnamefont
  {Y.}~\bibnamefont {Xia}}, \bibinfo {author} {\bibfnamefont {S.-Y.}\
  \bibnamefont {Xu}}, \bibinfo {author} {\bibfnamefont {D.}~\bibnamefont
  {Qian}}, \bibinfo {author} {\bibfnamefont {M.~Z.}\ \bibnamefont {Hasan}},
  \bibinfo {author} {\bibfnamefont {N.~P.}\ \bibnamefont {Ong}}, \bibinfo
  {author} {\bibfnamefont {A.}~\bibnamefont {Yazdani}}, \ and\ \bibinfo
  {author} {\bibfnamefont {R.~J.}\ \bibnamefont {Cava}},\ }\href {\doibase
  10.1103/PhysRevB.81.195203} {\bibfield  {journal} {\bibinfo  {journal} {Phys.
  Rev. B}\ }\textbf {\bibinfo {volume} {81}},\ \bibinfo {pages} {195203}
  (\bibinfo {year} {2010})}\BibitemShut {NoStop}%
\bibitem [{\citenamefont {Yang}\ \emph {et~al.}(2013)\citenamefont {Yang},
  \citenamefont {Dolev}, \citenamefont {Zhang}, \citenamefont {Zhao},
  \citenamefont {Fried}, \citenamefont {Schemm}, \citenamefont {Liu},
  \citenamefont {Palevski}, \citenamefont {Marshall}, \citenamefont {Risbud},\
  and\ \citenamefont {Kapitulnik}}]{bilayer2013}%
  \BibitemOpen
  \bibfield  {author} {\bibinfo {author} {\bibfnamefont {Q.~I.}\ \bibnamefont
  {Yang}}, \bibinfo {author} {\bibfnamefont {M.}~\bibnamefont {Dolev}},
  \bibinfo {author} {\bibfnamefont {L.}~\bibnamefont {Zhang}}, \bibinfo
  {author} {\bibfnamefont {J.}~\bibnamefont {Zhao}}, \bibinfo {author}
  {\bibfnamefont {A.~D.}\ \bibnamefont {Fried}}, \bibinfo {author}
  {\bibfnamefont {E.}~\bibnamefont {Schemm}}, \bibinfo {author} {\bibfnamefont
  {M.}~\bibnamefont {Liu}}, \bibinfo {author} {\bibfnamefont {A.}~\bibnamefont
  {Palevski}}, \bibinfo {author} {\bibfnamefont {A.~F.}\ \bibnamefont
  {Marshall}}, \bibinfo {author} {\bibfnamefont {S.~H.}\ \bibnamefont
  {Risbud}}, \ and\ \bibinfo {author} {\bibfnamefont {A.}~\bibnamefont
  {Kapitulnik}},\ }\href {\doibase 10.1103/PhysRevB.88.081407} {\bibfield
  {journal} {\bibinfo  {journal} {Phys. Rev. B}\ }\textbf {\bibinfo {volume}
  {88}},\ \bibinfo {pages} {081407} (\bibinfo {year} {2013})}\BibitemShut
  {NoStop}%
\bibitem [{\citenamefont {Kandala}\ \emph {et~al.}(2013)\citenamefont
  {Kandala}, \citenamefont {Richardella}, \citenamefont {Rench}, \citenamefont
  {Zhang}, \citenamefont {Flanagan},\ and\ \citenamefont
  {Samarth}}]{Samarth2013}%
  \BibitemOpen
  \bibfield  {author} {\bibinfo {author} {\bibfnamefont {A.}~\bibnamefont
  {Kandala}}, \bibinfo {author} {\bibfnamefont {A.}~\bibnamefont
  {Richardella}}, \bibinfo {author} {\bibfnamefont {D.~W.}\ \bibnamefont
  {Rench}}, \bibinfo {author} {\bibfnamefont {D.~M.}\ \bibnamefont {Zhang}},
  \bibinfo {author} {\bibfnamefont {T.~C.}\ \bibnamefont {Flanagan}}, \ and\
  \bibinfo {author} {\bibfnamefont {N.}~\bibnamefont {Samarth}},\ }\href
  {\doibase 10.1063/1.4831987} {\bibfield  {journal} {\bibinfo  {journal}
  {Appl. Phys. Lett.}\ }\textbf {\bibinfo {volume} {103}},\ \bibinfo {pages}
  {202409} (\bibinfo {year} {2013})}\BibitemShut {NoStop}%
\bibitem [{\citenamefont {Wei}\ \emph {et~al.}(2013)\citenamefont {Wei},
  \citenamefont {Katmis}, \citenamefont {Assaf}, \citenamefont {Steinberg},
  \citenamefont {Jarillo-Herrero}, \citenamefont {Heiman},\ and\ \citenamefont
  {Moodera}}]{Moodera2013}%
  \BibitemOpen
  \bibfield  {author} {\bibinfo {author} {\bibfnamefont {P.}~\bibnamefont
  {Wei}}, \bibinfo {author} {\bibfnamefont {F.}~\bibnamefont {Katmis}},
  \bibinfo {author} {\bibfnamefont {B.~A.}\ \bibnamefont {Assaf}}, \bibinfo
  {author} {\bibfnamefont {H.}~\bibnamefont {Steinberg}}, \bibinfo {author}
  {\bibfnamefont {P.}~\bibnamefont {Jarillo-Herrero}}, \bibinfo {author}
  {\bibfnamefont {D.}~\bibnamefont {Heiman}}, \ and\ \bibinfo {author}
  {\bibfnamefont {J.~S.}\ \bibnamefont {Moodera}},\ }\href {\doibase
  10.1103/PhysRevLett.110.186807} {\bibfield  {journal} {\bibinfo  {journal}
  {Phys. Rev. Lett.}\ }\textbf {\bibinfo {volume} {110}},\ \bibinfo {pages}
  {186807} (\bibinfo {year} {2013})}\BibitemShut {NoStop}%
\bibitem [{\citenamefont {Hirahara}\ \emph {et~al.}(2017)\citenamefont
  {Hirahara}, \citenamefont {Eremeev}, \citenamefont {Shirasawa}, \citenamefont
  {Okuyama}, \citenamefont {Kubo}, \citenamefont {Nakanishi}, \citenamefont
  {Akiyama}, \citenamefont {Takayama}, \citenamefont {Hajiri}, \citenamefont
  {Ideta}, \citenamefont {Matsunami}, \citenamefont {Sumida}, \citenamefont
  {Miyamoto}, \citenamefont {Takagi}, \citenamefont {Tanaka}, \citenamefont
  {Okuda}, \citenamefont {Yokoyama}, \citenamefont {Kimura}, \citenamefont
  {Hasegawa},\ and\ \citenamefont {Chulkov}}]{Chulkov2017}%
  \BibitemOpen
  \bibfield  {author} {\bibinfo {author} {\bibfnamefont {T.}~\bibnamefont
  {Hirahara}}, \bibinfo {author} {\bibfnamefont {S.~V.}\ \bibnamefont
  {Eremeev}}, \bibinfo {author} {\bibfnamefont {T.}~\bibnamefont {Shirasawa}},
  \bibinfo {author} {\bibfnamefont {Y.}~\bibnamefont {Okuyama}}, \bibinfo
  {author} {\bibfnamefont {T.}~\bibnamefont {Kubo}}, \bibinfo {author}
  {\bibfnamefont {R.}~\bibnamefont {Nakanishi}}, \bibinfo {author}
  {\bibfnamefont {R.}~\bibnamefont {Akiyama}}, \bibinfo {author} {\bibfnamefont
  {A.}~\bibnamefont {Takayama}}, \bibinfo {author} {\bibfnamefont
  {T.}~\bibnamefont {Hajiri}}, \bibinfo {author} {\bibfnamefont {S.-i.}\
  \bibnamefont {Ideta}}, \bibinfo {author} {\bibfnamefont {M.}~\bibnamefont
  {Matsunami}}, \bibinfo {author} {\bibfnamefont {K.}~\bibnamefont {Sumida}},
  \bibinfo {author} {\bibfnamefont {K.}~\bibnamefont {Miyamoto}}, \bibinfo
  {author} {\bibfnamefont {Y.}~\bibnamefont {Takagi}}, \bibinfo {author}
  {\bibfnamefont {K.}~\bibnamefont {Tanaka}}, \bibinfo {author} {\bibfnamefont
  {T.}~\bibnamefont {Okuda}}, \bibinfo {author} {\bibfnamefont
  {T.}~\bibnamefont {Yokoyama}}, \bibinfo {author} {\bibfnamefont {S.-i.}\
  \bibnamefont {Kimura}}, \bibinfo {author} {\bibfnamefont {S.}~\bibnamefont
  {Hasegawa}}, \ and\ \bibinfo {author} {\bibfnamefont {E.~V.}\ \bibnamefont
  {Chulkov}},\ }\href {\doibase 10.1021/acs.nanolett.7b00560} {\bibfield
  {journal} {\bibinfo  {journal} {Nano Lett.}\ }\textbf {\bibinfo {volume}
  {17}},\ \bibinfo {pages} {3493} (\bibinfo {year} {2017})}\BibitemShut
  {NoStop}%
\bibitem [{\citenamefont {Kou}\ \emph {et~al.}(2015)\citenamefont {Kou},
  \citenamefont {Pan}, \citenamefont {Wang}, \citenamefont {Fan}, \citenamefont
  {Choi}, \citenamefont {Lee}, \citenamefont {Nie}, \citenamefont {Murata},
  \citenamefont {Shao}, \citenamefont {Zhang},\ and\ \citenamefont
  {Wang}}]{Kou2015}%
  \BibitemOpen
  \bibfield  {author} {\bibinfo {author} {\bibfnamefont {X.}~\bibnamefont
  {Kou}}, \bibinfo {author} {\bibfnamefont {L.}~\bibnamefont {Pan}}, \bibinfo
  {author} {\bibfnamefont {J.}~\bibnamefont {Wang}}, \bibinfo {author}
  {\bibfnamefont {Y.}~\bibnamefont {Fan}}, \bibinfo {author} {\bibfnamefont
  {E.~S.}\ \bibnamefont {Choi}}, \bibinfo {author} {\bibfnamefont {W.-L.}\
  \bibnamefont {Lee}}, \bibinfo {author} {\bibfnamefont {T.}~\bibnamefont
  {Nie}}, \bibinfo {author} {\bibfnamefont {K.}~\bibnamefont {Murata}},
  \bibinfo {author} {\bibfnamefont {Q.}~\bibnamefont {Shao}}, \bibinfo {author}
  {\bibfnamefont {S.-C.}\ \bibnamefont {Zhang}}, \ and\ \bibinfo {author}
  {\bibfnamefont {K.~L.}\ \bibnamefont {Wang}},\ }\href {\doibase
  10.1038/ncomms9474} {\bibfield  {journal} {\bibinfo  {journal} {Nature
  Communications}\ }\textbf {\bibinfo {volume} {6}} (\bibinfo {year} {2015}),\
  10.1038/ncomms9474}\BibitemShut {NoStop}%
\bibitem [{\citenamefont {Katmis}\ \emph {et~al.}(2016)\citenamefont {Katmis},
  \citenamefont {Lauter}, \citenamefont {Nogueira}, \citenamefont {Assaf},
  \citenamefont {Jamer}, \citenamefont {Wei}, \citenamefont {Satpati},
  \citenamefont {Freeland}, \citenamefont {Eremin}, \citenamefont {Heiman},
  \citenamefont {Jarillo-Herrero},\ and\ \citenamefont
  {Moodera}}]{Moodera2016}%
  \BibitemOpen
  \bibfield  {author} {\bibinfo {author} {\bibfnamefont {F.}~\bibnamefont
  {Katmis}}, \bibinfo {author} {\bibfnamefont {V.}~\bibnamefont {Lauter}},
  \bibinfo {author} {\bibfnamefont {F.~S.}\ \bibnamefont {Nogueira}}, \bibinfo
  {author} {\bibfnamefont {B.~A.}\ \bibnamefont {Assaf}}, \bibinfo {author}
  {\bibfnamefont {M.~E.}\ \bibnamefont {Jamer}}, \bibinfo {author}
  {\bibfnamefont {P.}~\bibnamefont {Wei}}, \bibinfo {author} {\bibfnamefont
  {B.}~\bibnamefont {Satpati}}, \bibinfo {author} {\bibfnamefont {J.~W.}\
  \bibnamefont {Freeland}}, \bibinfo {author} {\bibfnamefont {I.}~\bibnamefont
  {Eremin}}, \bibinfo {author} {\bibfnamefont {D.}~\bibnamefont {Heiman}},
  \bibinfo {author} {\bibfnamefont {P.}~\bibnamefont {Jarillo-Herrero}}, \ and\
  \bibinfo {author} {\bibfnamefont {J.~S.}\ \bibnamefont {Moodera}},\ }\href
  {\doibase 10.1038/nature17635} {\bibfield  {journal} {\bibinfo  {journal}
  {Nat. Lett.}\ }\textbf {\bibinfo {volume} {533}},\ \bibinfo {pages} {513}
  (\bibinfo {year} {2016})}\BibitemShut {NoStop}%
\bibitem [{\citenamefont {Shapira}\ and\ \citenamefont
  {Reed}(1972)}]{EuS_ntype}%
  \BibitemOpen
  \bibfield  {author} {\bibinfo {author} {\bibfnamefont {Y.}~\bibnamefont
  {Shapira}}\ and\ \bibinfo {author} {\bibfnamefont {T.~B.}\ \bibnamefont
  {Reed}},\ }\href {\doibase 10.1103/PhysRevB.5.4877} {\bibfield  {journal}
  {\bibinfo  {journal} {Phys. Rev. B}\ }\textbf {\bibinfo {volume} {5}},\
  \bibinfo {pages} {4877} (\bibinfo {year} {1972})}\BibitemShut {NoStop}%
\bibitem [{\citenamefont {Keller}\ \emph {et~al.}(2002)\citenamefont {Keller},
  \citenamefont {Parker}, \citenamefont {Stankiewicz}, \citenamefont {Read},
  \citenamefont {Stampe}, \citenamefont {Kennedy}, \citenamefont {Xiong},\ and\
  \citenamefont {von Molnar}}]{Keller2002}%
  \BibitemOpen
  \bibfield  {author} {\bibinfo {author} {\bibfnamefont {J.}~\bibnamefont
  {Keller}}, \bibinfo {author} {\bibfnamefont {J.}~\bibnamefont {Parker}},
  \bibinfo {author} {\bibfnamefont {J.}~\bibnamefont {Stankiewicz}}, \bibinfo
  {author} {\bibfnamefont {D.}~\bibnamefont {Read}}, \bibinfo {author}
  {\bibfnamefont {P.}~\bibnamefont {Stampe}}, \bibinfo {author} {\bibfnamefont
  {R.}~\bibnamefont {Kennedy}}, \bibinfo {author} {\bibfnamefont
  {P.}~\bibnamefont {Xiong}}, \ and\ \bibinfo {author} {\bibfnamefont
  {S.}~\bibnamefont {von Molnar}},\ }\href {\doibase 10.1109/TMAG.2002.801977}
  {\bibfield  {journal} {\bibinfo  {journal} {IEEE Transactions on Magnetics}\
  }\textbf {\bibinfo {volume} {38}},\ \bibinfo {pages} {2673} (\bibinfo {year}
  {2002})},\ \bibinfo {note} {international Magnetics Conference (Intermag
  Europe 2002), Amsterdam, Netherlands, Apr 28-May 02, 2002}\BibitemShut
  {NoStop}%
\bibitem [{\citenamefont {Xia}\ \emph {et~al.}(2009)\citenamefont {Xia},
  \citenamefont {Qian}, \citenamefont {Hsieh}, \citenamefont {Wray},
  \citenamefont {Pal}, \citenamefont {Lin}, \citenamefont {Bansil},
  \citenamefont {Grauer}, \citenamefont {Hor}, \citenamefont {Cava},\ and\
  \citenamefont {Hasan}}]{Bi2Se3_ARPES1}%
  \BibitemOpen
  \bibfield  {author} {\bibinfo {author} {\bibfnamefont {Y.}~\bibnamefont
  {Xia}}, \bibinfo {author} {\bibfnamefont {D.}~\bibnamefont {Qian}}, \bibinfo
  {author} {\bibfnamefont {D.}~\bibnamefont {Hsieh}}, \bibinfo {author}
  {\bibfnamefont {L.}~\bibnamefont {Wray}}, \bibinfo {author} {\bibfnamefont
  {A.}~\bibnamefont {Pal}}, \bibinfo {author} {\bibfnamefont {H.}~\bibnamefont
  {Lin}}, \bibinfo {author} {\bibfnamefont {A.}~\bibnamefont {Bansil}},
  \bibinfo {author} {\bibfnamefont {D.}~\bibnamefont {Grauer}}, \bibinfo
  {author} {\bibfnamefont {Y.~S.}\ \bibnamefont {Hor}}, \bibinfo {author}
  {\bibfnamefont {R.~J.}\ \bibnamefont {Cava}}, \ and\ \bibinfo {author}
  {\bibfnamefont {M.~Z.}\ \bibnamefont {Hasan}},\ }\href {\doibase
  10.1038/nphys1274} {\bibfield  {journal} {\bibinfo  {journal} {Nat. Phys.}\
  }\textbf {\bibinfo {volume} {5}},\ \bibinfo {pages} {398} (\bibinfo {year}
  {2009})}\BibitemShut {NoStop}%
\bibitem [{\citenamefont {Zhang}\ \emph {et~al.}(2013)\citenamefont {Zhang},
  \citenamefont {Dolev}, \citenamefont {Yang}, \citenamefont {Hammond},
  \citenamefont {Zhou}, \citenamefont {Palevski}, \citenamefont {Chen},\ and\
  \citenamefont {Kapitulnik}}]{zhangli2013}%
  \BibitemOpen
  \bibfield  {author} {\bibinfo {author} {\bibfnamefont {L.}~\bibnamefont
  {Zhang}}, \bibinfo {author} {\bibfnamefont {M.}~\bibnamefont {Dolev}},
  \bibinfo {author} {\bibfnamefont {Q.~I.}\ \bibnamefont {Yang}}, \bibinfo
  {author} {\bibfnamefont {R.~H.}\ \bibnamefont {Hammond}}, \bibinfo {author}
  {\bibfnamefont {B.}~\bibnamefont {Zhou}}, \bibinfo {author} {\bibfnamefont
  {A.}~\bibnamefont {Palevski}}, \bibinfo {author} {\bibfnamefont
  {Y.}~\bibnamefont {Chen}}, \ and\ \bibinfo {author} {\bibfnamefont
  {A.}~\bibnamefont {Kapitulnik}},\ }\href {\doibase
  10.1103/PhysRevB.88.121103} {\bibfield  {journal} {\bibinfo  {journal} {Phys.
  Rev. B}\ }\textbf {\bibinfo {volume} {88}},\ \bibinfo {pages} {121103}
  (\bibinfo {year} {2013})}\BibitemShut {NoStop}%
\bibitem [{\citenamefont {Alpichshev}\ \emph {et~al.}(2012)\citenamefont
  {Alpichshev}, \citenamefont {Biswas}, \citenamefont {Balatsky}, \citenamefont
  {Analytis}, \citenamefont {Chu}, \citenamefont {Fisher},\ and\ \citenamefont
  {Kapitulnik}}]{Zhanybek3}%
  \BibitemOpen
  \bibfield  {author} {\bibinfo {author} {\bibfnamefont {Z.}~\bibnamefont
  {Alpichshev}}, \bibinfo {author} {\bibfnamefont {R.~R.}\ \bibnamefont
  {Biswas}}, \bibinfo {author} {\bibfnamefont {A.~V.}\ \bibnamefont
  {Balatsky}}, \bibinfo {author} {\bibfnamefont {J.~G.}\ \bibnamefont
  {Analytis}}, \bibinfo {author} {\bibfnamefont {J.-H.}\ \bibnamefont {Chu}},
  \bibinfo {author} {\bibfnamefont {I.~R.}\ \bibnamefont {Fisher}}, \ and\
  \bibinfo {author} {\bibfnamefont {A.}~\bibnamefont {Kapitulnik}},\ }\href
  {\doibase 10.1103/PhysRevLett.108.206402} {\bibfield  {journal} {\bibinfo
  {journal} {Phys. Rev. Lett.}\ }\textbf {\bibinfo {volume} {108}},\ \bibinfo
  {pages} {206402} (\bibinfo {year} {2012})}\BibitemShut {NoStop}%
\bibitem [{\citenamefont {Analytis}\ \emph {et~al.}(2010)\citenamefont
  {Analytis}, \citenamefont {Chu}, \citenamefont {Chen}, \citenamefont
  {Corredor}, \citenamefont {McDonald}, \citenamefont {Shen},\ and\
  \citenamefont {Fisher}}]{Fisher2010}%
  \BibitemOpen
  \bibfield  {author} {\bibinfo {author} {\bibfnamefont {J.~G.}\ \bibnamefont
  {Analytis}}, \bibinfo {author} {\bibfnamefont {J.-H.}\ \bibnamefont {Chu}},
  \bibinfo {author} {\bibfnamefont {Y.}~\bibnamefont {Chen}}, \bibinfo {author}
  {\bibfnamefont {F.}~\bibnamefont {Corredor}}, \bibinfo {author}
  {\bibfnamefont {R.~D.}\ \bibnamefont {McDonald}}, \bibinfo {author}
  {\bibfnamefont {Z.~X.}\ \bibnamefont {Shen}}, \ and\ \bibinfo {author}
  {\bibfnamefont {I.~R.}\ \bibnamefont {Fisher}},\ }\href {\doibase
  10.1103/PhysRevB.81.205407} {\bibfield  {journal} {\bibinfo  {journal} {Phys.
  Rev. B}\ }\textbf {\bibinfo {volume} {81}},\ \bibinfo {pages} {205407}
  (\bibinfo {year} {2010})}\BibitemShut {NoStop}%
\bibitem [{\citenamefont {Anderson}\ and\ \citenamefont
  {Krause}(1974)}]{SbStructure}%
  \BibitemOpen
  \bibfield  {author} {\bibinfo {author} {\bibfnamefont {T.~L.}\ \bibnamefont
  {Anderson}}\ and\ \bibinfo {author} {\bibfnamefont {H.~B.}\ \bibnamefont
  {Krause}},\ }\href {\doibase 10.1107/S0567740874004729} {\bibfield  {journal}
  {\bibinfo  {journal} {Acta Crystallogr., Sect. B: Struct. Sci.}\ }\textbf
  {\bibinfo {volume} {30}},\ \bibinfo {pages} {1307} (\bibinfo {year}
  {1974})}\BibitemShut {NoStop}%
\bibitem [{\citenamefont {Francombe}(1958)}]{BiStructure}%
  \BibitemOpen
  \bibfield  {author} {\bibinfo {author} {\bibfnamefont {M.~H.}\ \bibnamefont
  {Francombe}},\ }\href {http://stacks.iop.org/0508-3443/9/i=10/a=307}
  {\bibfield  {journal} {\bibinfo  {journal} {Br. J. Appl. Phys.}\ }\textbf
  {\bibinfo {volume} {9}},\ \bibinfo {pages} {415} (\bibinfo {year}
  {1958})}\BibitemShut {NoStop}%
\bibitem [{\citenamefont {Zhang}\ \emph {et~al.}(2009)\citenamefont {Zhang},
  \citenamefont {Liu}, \citenamefont {Qi}, \citenamefont {Dai}, \citenamefont
  {Fang},\ and\ \citenamefont {Zhang}}]{Zhang2009}%
  \BibitemOpen
  \bibfield  {author} {\bibinfo {author} {\bibfnamefont {H.}~\bibnamefont
  {Zhang}}, \bibinfo {author} {\bibfnamefont {C.-X.}\ \bibnamefont {Liu}},
  \bibinfo {author} {\bibfnamefont {X.-L.}\ \bibnamefont {Qi}}, \bibinfo
  {author} {\bibfnamefont {X.}~\bibnamefont {Dai}}, \bibinfo {author}
  {\bibfnamefont {Z.}~\bibnamefont {Fang}}, \ and\ \bibinfo {author}
  {\bibfnamefont {S.-C.}\ \bibnamefont {Zhang}},\ }\href {\doibase
  10.1038/NPHYS1270} {\bibfield  {journal} {\bibinfo  {journal} {Nature
  Physics}\ }\textbf {\bibinfo {volume} {5}},\ \bibinfo {pages} {438} (\bibinfo
  {year} {2009})}\BibitemShut {NoStop}%
\bibitem [{\citenamefont {Zhang}\ \emph {et~al.}(2011)\citenamefont {Zhang},
  \citenamefont {Chang}, \citenamefont {Zhang}, \citenamefont {Wen},
  \citenamefont {Feng}, \citenamefont {Li}, \citenamefont {Liu}, \citenamefont
  {He}, \citenamefont {Wang}, \citenamefont {Chen}, \citenamefont {Xue},
  \citenamefont {Ma},\ and\ \citenamefont {Wang}}]{ZhangJS2011}%
  \BibitemOpen
  \bibfield  {author} {\bibinfo {author} {\bibfnamefont {J.}~\bibnamefont
  {Zhang}}, \bibinfo {author} {\bibfnamefont {C.-Z.}\ \bibnamefont {Chang}},
  \bibinfo {author} {\bibfnamefont {Z.}~\bibnamefont {Zhang}}, \bibinfo
  {author} {\bibfnamefont {J.}~\bibnamefont {Wen}}, \bibinfo {author}
  {\bibfnamefont {X.}~\bibnamefont {Feng}}, \bibinfo {author} {\bibfnamefont
  {K.}~\bibnamefont {Li}}, \bibinfo {author} {\bibfnamefont {M.}~\bibnamefont
  {Liu}}, \bibinfo {author} {\bibfnamefont {K.}~\bibnamefont {He}}, \bibinfo
  {author} {\bibfnamefont {L.}~\bibnamefont {Wang}}, \bibinfo {author}
  {\bibfnamefont {X.}~\bibnamefont {Chen}}, \bibinfo {author} {\bibfnamefont
  {Q.-K.}\ \bibnamefont {Xue}}, \bibinfo {author} {\bibfnamefont
  {X.}~\bibnamefont {Ma}}, \ and\ \bibinfo {author} {\bibfnamefont
  {Y.}~\bibnamefont {Wang}},\ }\href {\doibase 10.1038/ncomms1588} {\bibfield
  {journal} {\bibinfo  {journal} {Nat. Commun.}\ ,\ \bibinfo {pages} {574}}
  (\bibinfo {year} {2011})}\BibitemShut {NoStop}%
\bibitem [{\citenamefont {Kou}\ \emph {et~al.}(2014)\citenamefont {Kou},
  \citenamefont {Guo}, \citenamefont {Fan}, \citenamefont {Pan}, \citenamefont
  {Lang}, \citenamefont {Jiang}, \citenamefont {Shao}, \citenamefont {Nie},
  \citenamefont {Murata}, \citenamefont {Tang}, \citenamefont {Wang},
  \citenamefont {He}, \citenamefont {Lee}, \citenamefont {Lee},\ and\
  \citenamefont {Wang}}]{Kou2014}%
  \BibitemOpen
  \bibfield  {author} {\bibinfo {author} {\bibfnamefont {X.}~\bibnamefont
  {Kou}}, \bibinfo {author} {\bibfnamefont {S.-T.}\ \bibnamefont {Guo}},
  \bibinfo {author} {\bibfnamefont {Y.}~\bibnamefont {Fan}}, \bibinfo {author}
  {\bibfnamefont {L.}~\bibnamefont {Pan}}, \bibinfo {author} {\bibfnamefont
  {M.}~\bibnamefont {Lang}}, \bibinfo {author} {\bibfnamefont {Y.}~\bibnamefont
  {Jiang}}, \bibinfo {author} {\bibfnamefont {Q.}~\bibnamefont {Shao}},
  \bibinfo {author} {\bibfnamefont {T.}~\bibnamefont {Nie}}, \bibinfo {author}
  {\bibfnamefont {K.}~\bibnamefont {Murata}}, \bibinfo {author} {\bibfnamefont
  {J.}~\bibnamefont {Tang}}, \bibinfo {author} {\bibfnamefont {Y.}~\bibnamefont
  {Wang}}, \bibinfo {author} {\bibfnamefont {L.}~\bibnamefont {He}}, \bibinfo
  {author} {\bibfnamefont {T.-K.}\ \bibnamefont {Lee}}, \bibinfo {author}
  {\bibfnamefont {W.-L.}\ \bibnamefont {Lee}}, \ and\ \bibinfo {author}
  {\bibfnamefont {K.~L.}\ \bibnamefont {Wang}},\ }\href {\doibase
  10.1103/PhysRevLett.113.137201} {\bibfield  {journal} {\bibinfo  {journal}
  {Phys. Rev. Lett.}\ }\textbf {\bibinfo {volume} {113}},\ \bibinfo {pages}
  {137201} (\bibinfo {year} {2014})}\BibitemShut {NoStop}%
\bibitem [{\citenamefont {Yang}\ \emph {et~al.}(2014)\citenamefont {Yang},
  \citenamefont {Zhao}, \citenamefont {Zhang}, \citenamefont {Dolev},
  \citenamefont {Fried}, \citenamefont {Marshall}, \citenamefont {Risbud},\
  and\ \citenamefont {Kapitulnik}}]{EuS_PLD}%
  \BibitemOpen
  \bibfield  {author} {\bibinfo {author} {\bibfnamefont {Q.~I.}\ \bibnamefont
  {Yang}}, \bibinfo {author} {\bibfnamefont {J.}~\bibnamefont {Zhao}}, \bibinfo
  {author} {\bibfnamefont {L.}~\bibnamefont {Zhang}}, \bibinfo {author}
  {\bibfnamefont {M.}~\bibnamefont {Dolev}}, \bibinfo {author} {\bibfnamefont
  {A.~D.}\ \bibnamefont {Fried}}, \bibinfo {author} {\bibfnamefont {A.~F.}\
  \bibnamefont {Marshall}}, \bibinfo {author} {\bibfnamefont {S.~H.}\
  \bibnamefont {Risbud}}, \ and\ \bibinfo {author} {\bibfnamefont
  {A.}~\bibnamefont {Kapitulnik}},\ }\href {\doibase 10.1063/1.4866265}
  {\bibfield  {journal} {\bibinfo  {journal} {Appl. Phys. Lett.}\ }\textbf
  {\bibinfo {volume} {104}},\ \bibinfo {pages} {082402} (\bibinfo {year}
  {2014})}\BibitemShut {NoStop}%
\bibitem [{\citenamefont {Obara}\ \emph {et~al.}(2009)\citenamefont {Obara},
  \citenamefont {Higomo}, \citenamefont {Ohta}, \citenamefont {Yamamoto},
  \citenamefont {Ueno},\ and\ \citenamefont {Iida}}]{telluride_PLD1}%
  \BibitemOpen
  \bibfield  {author} {\bibinfo {author} {\bibfnamefont {H.}~\bibnamefont
  {Obara}}, \bibinfo {author} {\bibfnamefont {S.}~\bibnamefont {Higomo}},
  \bibinfo {author} {\bibfnamefont {M.}~\bibnamefont {Ohta}}, \bibinfo {author}
  {\bibfnamefont {A.}~\bibnamefont {Yamamoto}}, \bibinfo {author}
  {\bibfnamefont {K.}~\bibnamefont {Ueno}}, \ and\ \bibinfo {author}
  {\bibfnamefont {T.}~\bibnamefont {Iida}},\ }\href
  {http://stacks.iop.org/1347-4065/48/i=8R/a=085506} {\bibfield  {journal}
  {\bibinfo  {journal} {Jpn. J. Appl. Phys.}\ }\textbf {\bibinfo {volume}
  {48}},\ \bibinfo {pages} {085506} (\bibinfo {year} {2009})}\BibitemShut
  {NoStop}%
\bibitem [{\citenamefont {Shaik}\ and\ \citenamefont
  {Motaleb}(2013)}]{telluride_PLD2}%
  \BibitemOpen
  \bibfield  {author} {\bibinfo {author} {\bibfnamefont {M.}~\bibnamefont
  {Shaik}}\ and\ \bibinfo {author} {\bibfnamefont {I.~A.}\ \bibnamefont
  {Motaleb}},\ }in\ \href {\doibase 10.1109/EIT.2013.6632706} {\emph {\bibinfo
  {booktitle} {IEEE International Conference on Electro-Information
  Technology}}}\ (\bibinfo {year} {2013})\ pp.\ \bibinfo {pages}
  {1--6}\BibitemShut {NoStop}%
\bibitem [{\citenamefont {Korzenski}\ \emph {et~al.}(2001)\citenamefont
  {Korzenski}, \citenamefont {Lecoeur}, \citenamefont {Mercey}, \citenamefont
  {Camy},\ and\ \citenamefont {Doualan}}]{PLD_alt_target2}%
  \BibitemOpen
  \bibfield  {author} {\bibinfo {author} {\bibfnamefont {M.~B.}\ \bibnamefont
  {Korzenski}}, \bibinfo {author} {\bibfnamefont {P.}~\bibnamefont {Lecoeur}},
  \bibinfo {author} {\bibfnamefont {B.}~\bibnamefont {Mercey}}, \bibinfo
  {author} {\bibfnamefont {P.}~\bibnamefont {Camy}}, \ and\ \bibinfo {author}
  {\bibfnamefont {J.-L.}\ \bibnamefont {Doualan}},\ }\href {\doibase
  10.1063/1.1347026} {\bibfield  {journal} {\bibinfo  {journal} {Appl. Phys.
  Lett.}\ }\textbf {\bibinfo {volume} {78}},\ \bibinfo {pages} {1210} (\bibinfo
  {year} {2001})}\BibitemShut {NoStop}%
\bibitem [{\citenamefont {Ambrosini}\ and\ \citenamefont
  {Hamet}(2003)}]{PLD_alt_target}%
  \BibitemOpen
  \bibfield  {author} {\bibinfo {author} {\bibfnamefont {A.}~\bibnamefont
  {Ambrosini}}\ and\ \bibinfo {author} {\bibfnamefont {J.-F.}\ \bibnamefont
  {Hamet}},\ }\href {\doibase 10.1063/1.1541116} {\bibfield  {journal}
  {\bibinfo  {journal} {Appl. Phys. Lett.}\ }\textbf {\bibinfo {volume} {82}},\
  \bibinfo {pages} {727} (\bibinfo {year} {2003})}\BibitemShut {NoStop}%
\bibitem [{Note1()}]{Note1}%
  \BibitemOpen
  \bibinfo {note} {See supplemental material at [URL will be inserted by
  publisher] for details of sample fabrication and basic characterization,
  which includes refs.~\protect \rev@citealpnum {EuS_PLD, telluride_PLD1,
  telluride_PLD2, PLD_alt_target2, PLD_alt_target, ZhangJS2011,
  SbStructure}.}\BibitemShut {Stop}%
\bibitem [{\citenamefont {Casañ-Pastor}\ \emph {et~al.}(1991)\citenamefont
  {Casañ-Pastor}, \citenamefont {Gomez-Romero},\ and\ \citenamefont
  {Baker}}]{squid_center_error}%
  \BibitemOpen
  \bibfield  {author} {\bibinfo {author} {\bibfnamefont {N.}~\bibnamefont
  {Casañ-Pastor}}, \bibinfo {author} {\bibfnamefont {P.}~\bibnamefont
  {Gomez-Romero}}, \ and\ \bibinfo {author} {\bibfnamefont {L.~C.}\
  \bibnamefont {Baker}},\ }\href {\doibase 10.1063/1.348132} {\bibfield
  {journal} {\bibinfo  {journal} {J. Appl. Phys.}\ }\textbf {\bibinfo {volume}
  {69}},\ \bibinfo {pages} {5088} (\bibinfo {year} {1991})}\BibitemShut
  {NoStop}%
\bibitem [{\citenamefont {Chiarelli}\ \emph {et~al.}(1993)\citenamefont
  {Chiarelli}, \citenamefont {Novak}, \citenamefont {Rassat},\ and\
  \citenamefont {Tholence}}]{ac_nitroxide}%
  \BibitemOpen
  \bibfield  {author} {\bibinfo {author} {\bibfnamefont {R.}~\bibnamefont
  {Chiarelli}}, \bibinfo {author} {\bibfnamefont {M.~A.}\ \bibnamefont
  {Novak}}, \bibinfo {author} {\bibfnamefont {A.}~\bibnamefont {Rassat}}, \
  and\ \bibinfo {author} {\bibfnamefont {J.~L.}\ \bibnamefont {Tholence}},\
  }\href@noop {} {\bibfield  {journal} {\bibinfo  {journal} {Nature}\ }\textbf
  {\bibinfo {volume} {363}},\ \bibinfo {pages} {147} (\bibinfo {year}
  {1993})}\BibitemShut {NoStop}%
\bibitem [{\citenamefont {Sarkissian}(1981)}]{ac_spin_glass}%
  \BibitemOpen
  \bibfield  {author} {\bibinfo {author} {\bibfnamefont {B.~V.~B.}\
  \bibnamefont {Sarkissian}},\ }\href
  {http://stacks.iop.org/0305-4608/11/i=10/a=029} {\bibfield  {journal}
  {\bibinfo  {journal} {J. Phys. F: Met. Phys.}\ }\textbf {\bibinfo {volume}
  {11}},\ \bibinfo {pages} {2191} (\bibinfo {year} {1981})}\BibitemShut
  {NoStop}%
\bibitem [{\citenamefont {Dormann}\ \emph {et~al.}(1988)\citenamefont
  {Dormann}, \citenamefont {Bessais},\ and\ \citenamefont
  {Fiorani}}]{ac_superpara}%
  \BibitemOpen
  \bibfield  {author} {\bibinfo {author} {\bibfnamefont {J.~L.}\ \bibnamefont
  {Dormann}}, \bibinfo {author} {\bibfnamefont {L.}~\bibnamefont {Bessais}}, \
  and\ \bibinfo {author} {\bibfnamefont {D.}~\bibnamefont {Fiorani}},\ }\href
  {http://stacks.iop.org/0022-3719/21/i=10/a=019} {\bibfield  {journal}
  {\bibinfo  {journal} {J. Phys. C: Solid State Phys.}\ }\textbf {\bibinfo
  {volume} {21}},\ \bibinfo {pages} {2015} (\bibinfo {year}
  {1988})}\BibitemShut {NoStop}%
\bibitem [{\citenamefont {Ando}\ \emph {et~al.}(1994)\citenamefont {Ando},
  \citenamefont {Kubota}, \citenamefont {Sato},\ and\ \citenamefont
  {Terasaki}}]{Ando1994}%
  \BibitemOpen
  \bibfield  {author} {\bibinfo {author} {\bibfnamefont {Y.}~\bibnamefont
  {Ando}}, \bibinfo {author} {\bibfnamefont {H.}~\bibnamefont {Kubota}},
  \bibinfo {author} {\bibfnamefont {Y.}~\bibnamefont {Sato}}, \ and\ \bibinfo
  {author} {\bibfnamefont {I.}~\bibnamefont {Terasaki}},\ }\href {\doibase
  10.1103/PhysRevB.50.9680} {\bibfield  {journal} {\bibinfo  {journal} {Phys.
  Rev. B}\ }\textbf {\bibinfo {volume} {50}},\ \bibinfo {pages} {9680}
  (\bibinfo {year} {1994})}\BibitemShut {NoStop}%
\bibitem [{\citenamefont {Gegenwart}\ \emph {et~al.}(2005)\citenamefont
  {Gegenwart}, \citenamefont {Custers}, \citenamefont {Tokiwa}, \citenamefont
  {Geibel},\ and\ \citenamefont {Steglich}}]{Gegenwart2005}%
  \BibitemOpen
  \bibfield  {author} {\bibinfo {author} {\bibfnamefont {P.}~\bibnamefont
  {Gegenwart}}, \bibinfo {author} {\bibfnamefont {J.}~\bibnamefont {Custers}},
  \bibinfo {author} {\bibfnamefont {Y.}~\bibnamefont {Tokiwa}}, \bibinfo
  {author} {\bibfnamefont {C.}~\bibnamefont {Geibel}}, \ and\ \bibinfo {author}
  {\bibfnamefont {F.}~\bibnamefont {Steglich}},\ }\href {\doibase
  10.1103/PhysRevLett.94.076402} {\bibfield  {journal} {\bibinfo  {journal}
  {Phys. Rev. Lett.}\ }\textbf {\bibinfo {volume} {94}},\ \bibinfo {pages}
  {076402} (\bibinfo {year} {2005})}\BibitemShut {NoStop}%
\bibitem [{\citenamefont {Schemm}\ \emph {et~al.}(2014)\citenamefont {Schemm},
  \citenamefont {Gannon}, \citenamefont {Wishne}, \citenamefont {Halperin},\
  and\ \citenamefont {Kapitulnik}}]{Schemm2014}%
  \BibitemOpen
  \bibfield  {author} {\bibinfo {author} {\bibfnamefont {E.~R.}\ \bibnamefont
  {Schemm}}, \bibinfo {author} {\bibfnamefont {W.~J.}\ \bibnamefont {Gannon}},
  \bibinfo {author} {\bibfnamefont {C.~M.}\ \bibnamefont {Wishne}}, \bibinfo
  {author} {\bibfnamefont {W.~P.}\ \bibnamefont {Halperin}}, \ and\ \bibinfo
  {author} {\bibfnamefont {A.}~\bibnamefont {Kapitulnik}},\ }\href {\doibase
  10.1126/science.1248552} {\bibfield  {journal} {\bibinfo  {journal}
  {Science}\ }\textbf {\bibinfo {volume} {345}},\ \bibinfo {pages} {190}
  (\bibinfo {year} {2014})}\BibitemShut {NoStop}%
\bibitem [{\citenamefont {Jeanneret}\ \emph {et~al.}(1989)\citenamefont
  {Jeanneret}, \citenamefont {Gavilano}, \citenamefont {Racine}, \citenamefont
  {Leemann},\ and\ \citenamefont {Martinoli}}]{Jeanneret1989}%
  \BibitemOpen
  \bibfield  {author} {\bibinfo {author} {\bibfnamefont {B.}~\bibnamefont
  {Jeanneret}}, \bibinfo {author} {\bibfnamefont {J.~L.}\ \bibnamefont
  {Gavilano}}, \bibinfo {author} {\bibfnamefont {G.~A.}\ \bibnamefont
  {Racine}}, \bibinfo {author} {\bibfnamefont {C.}~\bibnamefont {Leemann}}, \
  and\ \bibinfo {author} {\bibfnamefont {P.}~\bibnamefont {Martinoli}},\ }\href
  {\doibase 10.1063/1.102053} {\bibfield  {journal} {\bibinfo  {journal} {Appl.
  Phys. Lett.}\ }\textbf {\bibinfo {volume} {55}},\ \bibinfo {pages} {2336}
  (\bibinfo {year} {1989})}\BibitemShut {NoStop}%
\bibitem [{\citenamefont {Yazdani}\ \emph {et~al.}(1993)\citenamefont
  {Yazdani}, \citenamefont {White}, \citenamefont {Hahn}, \citenamefont
  {Gabay}, \citenamefont {Beasley},\ and\ \citenamefont
  {Kapitulnik}}]{Yazdani1993}%
  \BibitemOpen
  \bibfield  {author} {\bibinfo {author} {\bibfnamefont {A.}~\bibnamefont
  {Yazdani}}, \bibinfo {author} {\bibfnamefont {W.~R.}\ \bibnamefont {White}},
  \bibinfo {author} {\bibfnamefont {M.~R.}\ \bibnamefont {Hahn}}, \bibinfo
  {author} {\bibfnamefont {M.}~\bibnamefont {Gabay}}, \bibinfo {author}
  {\bibfnamefont {M.~R.}\ \bibnamefont {Beasley}}, \ and\ \bibinfo {author}
  {\bibfnamefont {A.}~\bibnamefont {Kapitulnik}},\ }\href {\doibase
  10.1103/PhysRevLett.70.505} {\bibfield  {journal} {\bibinfo  {journal} {Phys.
  Rev. Lett.}\ }\textbf {\bibinfo {volume} {70}},\ \bibinfo {pages} {505}
  (\bibinfo {year} {1993})}\BibitemShut {NoStop}%
\bibitem [{\citenamefont {Yazdani}(1994)}]{YazdaniThesis}%
  \BibitemOpen
  \bibfield  {author} {\bibinfo {author} {\bibfnamefont {A.}~\bibnamefont
  {Yazdani}},\ }\emph {\bibinfo {title} {Phase Transitions in Two-Dimensional
  Superconductors}},\ \href@noop {} {Ph.D. thesis},\ \bibinfo  {school}
  {Stanford University}, \bibinfo {address} {Stanford, CA 94305} (\bibinfo
  {year} {1994})\BibitemShut {NoStop}%
\bibitem [{\citenamefont {Dunlavy}\ and\ \citenamefont
  {Venus}(2004)}]{Venus2004}%
  \BibitemOpen
  \bibfield  {author} {\bibinfo {author} {\bibfnamefont {M.~J.}\ \bibnamefont
  {Dunlavy}}\ and\ \bibinfo {author} {\bibfnamefont {D.}~\bibnamefont
  {Venus}},\ }\href {\doibase 10.1103/PhysRevB.69.094411} {\bibfield  {journal}
  {\bibinfo  {journal} {Phys. Rev. B}\ }\textbf {\bibinfo {volume} {69}},\
  \bibinfo {pages} {094411} (\bibinfo {year} {2004})}\BibitemShut {NoStop}%
\bibitem [{\citenamefont {Jensen}\ \emph {et~al.}(2003)\citenamefont {Jensen},
  \citenamefont {Knappmann}, \citenamefont {Wulfhekel},\ and\ \citenamefont
  {Oepen}}]{Jensen2003}%
  \BibitemOpen
  \bibfield  {author} {\bibinfo {author} {\bibfnamefont {P.~J.}\ \bibnamefont
  {Jensen}}, \bibinfo {author} {\bibfnamefont {S.}~\bibnamefont {Knappmann}},
  \bibinfo {author} {\bibfnamefont {W.}~\bibnamefont {Wulfhekel}}, \ and\
  \bibinfo {author} {\bibfnamefont {H.~P.}\ \bibnamefont {Oepen}},\ }\href
  {\doibase 10.1103/PhysRevB.67.184417} {\bibfield  {journal} {\bibinfo
  {journal} {Phys. Rev. B}\ }\textbf {\bibinfo {volume} {67}},\ \bibinfo
  {pages} {184417} (\bibinfo {year} {2003})}\BibitemShut {NoStop}%
\bibitem [{\citenamefont {Cho}(1967)}]{EuS_band1}%
  \BibitemOpen
  \bibfield  {author} {\bibinfo {author} {\bibfnamefont {S.~J.}\ \bibnamefont
  {Cho}},\ }\href {\doibase 10.1103/PhysRev.157.632} {\bibfield  {journal}
  {\bibinfo  {journal} {Phys. Rev.}\ }\textbf {\bibinfo {volume} {157}},\
  \bibinfo {pages} {632} (\bibinfo {year} {1967})}\BibitemShut {NoStop}%
\bibitem [{\citenamefont {M\"uller}\ and\ \citenamefont
  {Nolting}(2002)}]{EuS_band2}%
  \BibitemOpen
  \bibfield  {author} {\bibinfo {author} {\bibfnamefont {W.}~\bibnamefont
  {M\"uller}}\ and\ \bibinfo {author} {\bibfnamefont {W.}~\bibnamefont
  {Nolting}},\ }\href {\doibase 10.1103/PhysRevB.66.085205} {\bibfield
  {journal} {\bibinfo  {journal} {Phys. Rev. B}\ }\textbf {\bibinfo {volume}
  {66}},\ \bibinfo {pages} {085205} (\bibinfo {year} {2002})}\BibitemShut
  {NoStop}%
\bibitem [{\citenamefont {Kim}\ \emph {et~al.}(2011)\citenamefont {Kim},
  \citenamefont {Brahlek}, \citenamefont {Bansal}, \citenamefont {Edrey},
  \citenamefont {Kapilevich}, \citenamefont {Iida}, \citenamefont {Tanimura},
  \citenamefont {Horibe}, \citenamefont {Cheong},\ and\ \citenamefont
  {Oh}}]{TI_WAL_thickness}%
  \BibitemOpen
  \bibfield  {author} {\bibinfo {author} {\bibfnamefont {Y.~S.}\ \bibnamefont
  {Kim}}, \bibinfo {author} {\bibfnamefont {M.}~\bibnamefont {Brahlek}},
  \bibinfo {author} {\bibfnamefont {N.}~\bibnamefont {Bansal}}, \bibinfo
  {author} {\bibfnamefont {E.}~\bibnamefont {Edrey}}, \bibinfo {author}
  {\bibfnamefont {G.~A.}\ \bibnamefont {Kapilevich}}, \bibinfo {author}
  {\bibfnamefont {K.}~\bibnamefont {Iida}}, \bibinfo {author} {\bibfnamefont
  {M.}~\bibnamefont {Tanimura}}, \bibinfo {author} {\bibfnamefont
  {Y.}~\bibnamefont {Horibe}}, \bibinfo {author} {\bibfnamefont {S.-W.}\
  \bibnamefont {Cheong}}, \ and\ \bibinfo {author} {\bibfnamefont
  {S.}~\bibnamefont {Oh}},\ }\href {\doibase 10.1103/PhysRevB.84.073109}
  {\bibfield  {journal} {\bibinfo  {journal} {Phys. Rev. B}\ }\textbf {\bibinfo
  {volume} {84}},\ \bibinfo {pages} {073109} (\bibinfo {year}
  {2011})}\BibitemShut {NoStop}%
\bibitem [{\citenamefont {Garate}\ and\ \citenamefont
  {Glazman}(2012)}]{Glazman2012}%
  \BibitemOpen
  \bibfield  {author} {\bibinfo {author} {\bibfnamefont {I.}~\bibnamefont
  {Garate}}\ and\ \bibinfo {author} {\bibfnamefont {L.}~\bibnamefont
  {Glazman}},\ }\href {\doibase 10.1103/PhysRevB.86.035422} {\bibfield
  {journal} {\bibinfo  {journal} {Phys. Rev. B}\ }\textbf {\bibinfo {volume}
  {86}},\ \bibinfo {pages} {035422} (\bibinfo {year} {2012})}\BibitemShut
  {NoStop}%
\bibitem [{\citenamefont {Lu}\ \emph {et~al.}(2011)\citenamefont {Lu},
  \citenamefont {Shi},\ and\ \citenamefont {Shen}}]{Lu2011}%
  \BibitemOpen
  \bibfield  {author} {\bibinfo {author} {\bibfnamefont {H.-Z.}\ \bibnamefont
  {Lu}}, \bibinfo {author} {\bibfnamefont {J.}~\bibnamefont {Shi}}, \ and\
  \bibinfo {author} {\bibfnamefont {S.-Q.}\ \bibnamefont {Shen}},\ }\href
  {\doibase 10.1103/PhysRevLett.107.076801} {\bibfield  {journal} {\bibinfo
  {journal} {Phys. Rev. Lett.}\ }\textbf {\bibinfo {volume} {107}},\ \bibinfo
  {pages} {076801} (\bibinfo {year} {2011})}\BibitemShut {NoStop}%
\bibitem [{\citenamefont {Dugaev}\ \emph {et~al.}(2001)\citenamefont {Dugaev},
  \citenamefont {Bruno},\ and\ \citenamefont {Barna\ifmmode~\acute{s}\else
  \'{s}\fi{}}}]{WL_ferromagnetism}%
  \BibitemOpen
  \bibfield  {author} {\bibinfo {author} {\bibfnamefont {V.~K.}\ \bibnamefont
  {Dugaev}}, \bibinfo {author} {\bibfnamefont {P.}~\bibnamefont {Bruno}}, \
  and\ \bibinfo {author} {\bibfnamefont {J.}~\bibnamefont
  {Barna\ifmmode~\acute{s}\else \'{s}\fi{}}},\ }\href {\doibase
  10.1103/PhysRevB.64.144423} {\bibfield  {journal} {\bibinfo  {journal} {Phys.
  Rev. B}\ }\textbf {\bibinfo {volume} {64}},\ \bibinfo {pages} {144423}
  (\bibinfo {year} {2001})}\BibitemShut {NoStop}%
\bibitem [{\citenamefont {Lee}\ \emph {et~al.}(2017)\citenamefont {Lee},
  \citenamefont {Richardella}, \citenamefont {Fraleigh}, \citenamefont {Liu},
  \citenamefont {Zhao},\ and\ \citenamefont {Samarth}}]{Samarth2017}%
  \BibitemOpen
  \bibfield  {author} {\bibinfo {author} {\bibfnamefont {J.~S.}\ \bibnamefont
  {Lee}}, \bibinfo {author} {\bibfnamefont {A.}~\bibnamefont {Richardella}},
  \bibinfo {author} {\bibfnamefont {R.~D.}\ \bibnamefont {Fraleigh}}, \bibinfo
  {author} {\bibfnamefont {C.-X.}\ \bibnamefont {Liu}}, \bibinfo {author}
  {\bibfnamefont {W.}~\bibnamefont {Zhao}}, \ and\ \bibinfo {author}
  {\bibfnamefont {N.}~\bibnamefont {Samarth}},\ }\href
  {https://arxiv.org/abs/1706.04661} {} (\bibinfo {year} {2017}),\ \Eprint
  {http://arxiv.org/abs/1706.04661} {arXiv:1706.04661 [cond-mat.mes-hall]}
  \BibitemShut {NoStop}%
\bibitem [{\citenamefont {Zheng}\ \emph {et~al.}(2016)\citenamefont {Zheng},
  \citenamefont {Wang}, \citenamefont {Yang}, \citenamefont {Wang},
  \citenamefont {Du}, \citenamefont {Ning}, \citenamefont {Yang}, \citenamefont
  {Lu}, \citenamefont {Zhang},\ and\ \citenamefont {Tian}}]{Tian2016}%
  \BibitemOpen
  \bibfield  {author} {\bibinfo {author} {\bibfnamefont {G.}~\bibnamefont
  {Zheng}}, \bibinfo {author} {\bibfnamefont {N.}~\bibnamefont {Wang}},
  \bibinfo {author} {\bibfnamefont {J.}~\bibnamefont {Yang}}, \bibinfo {author}
  {\bibfnamefont {W.}~\bibnamefont {Wang}}, \bibinfo {author} {\bibfnamefont
  {H.}~\bibnamefont {Du}}, \bibinfo {author} {\bibfnamefont {W.}~\bibnamefont
  {Ning}}, \bibinfo {author} {\bibfnamefont {Z.}~\bibnamefont {Yang}}, \bibinfo
  {author} {\bibfnamefont {H.-Z.}\ \bibnamefont {Lu}}, \bibinfo {author}
  {\bibfnamefont {Y.}~\bibnamefont {Zhang}}, \ and\ \bibinfo {author}
  {\bibfnamefont {M.}~\bibnamefont {Tian}},\ }\href {\doibase
  10.1038/srep21334} {\bibfield  {journal} {\bibinfo  {journal} {Sci. Rep.}\
  }\textbf {\bibinfo {volume} {6}},\ \bibinfo {pages} {21134} (\bibinfo {year}
  {2016})}\BibitemShut {NoStop}%
\bibitem [{\citenamefont {Mott}\ and\ \citenamefont {Davis}(2012)}]{Mott_book}%
  \BibitemOpen
  \bibfield  {author} {\bibinfo {author} {\bibfnamefont {N.~F.}\ \bibnamefont
  {Mott}}\ and\ \bibinfo {author} {\bibfnamefont {E.~A.}\ \bibnamefont
  {Davis}},\ }\href@noop {} {\emph {\bibinfo {title} {Electronic Processes in
  Non-Crystalline Materials}}},\ \bibinfo {edition} {2nd}\ ed.\ (\bibinfo
  {publisher} {Oxford University Press},\ \bibinfo {year} {2012})\BibitemShut
  {NoStop}%
\bibitem [{\citenamefont {Fradkin}(1986)}]{Fradkin1986b}%
  \BibitemOpen
  \bibfield  {author} {\bibinfo {author} {\bibfnamefont {E.}~\bibnamefont
  {Fradkin}},\ }\href {\doibase 10.1103/PhysRevB.33.3263} {\bibfield  {journal}
  {\bibinfo  {journal} {Phys. Rev. B}\ }\textbf {\bibinfo {volume} {33}},\
  \bibinfo {pages} {3263} (\bibinfo {year} {1986})}\BibitemShut {NoStop}%
\bibitem [{\citenamefont {Jiang}\ \emph {et~al.}(2014)\citenamefont {Jiang},
  \citenamefont {Katmis}, \citenamefont {Tang}, \citenamefont {Wei},
  \citenamefont {Moodera},\ and\ \citenamefont {Shi}}]{Shi2014}%
  \BibitemOpen
  \bibfield  {author} {\bibinfo {author} {\bibfnamefont {Z.}~\bibnamefont
  {Jiang}}, \bibinfo {author} {\bibfnamefont {F.}~\bibnamefont {Katmis}},
  \bibinfo {author} {\bibfnamefont {C.}~\bibnamefont {Tang}}, \bibinfo {author}
  {\bibfnamefont {P.}~\bibnamefont {Wei}}, \bibinfo {author} {\bibfnamefont
  {J.~S.}\ \bibnamefont {Moodera}}, \ and\ \bibinfo {author} {\bibfnamefont
  {J.}~\bibnamefont {Shi}},\ }\href {\doibase 10.1063/1.4881975} {\bibfield
  {journal} {\bibinfo  {journal} {Appl. Phys. Lett.}\ }\textbf {\bibinfo
  {volume} {104}},\ \bibinfo {pages} {222409} (\bibinfo {year}
  {2014})}\BibitemShut {NoStop}%
\bibitem [{\citenamefont {Alegria}\ \emph {et~al.}(2014)\citenamefont
  {Alegria}, \citenamefont {Ji}, \citenamefont {Yao}, \citenamefont {Clarke},
  \citenamefont {Cava},\ and\ \citenamefont {Petta}}]{Petta2014}%
  \BibitemOpen
  \bibfield  {author} {\bibinfo {author} {\bibfnamefont {L.~D.}\ \bibnamefont
  {Alegria}}, \bibinfo {author} {\bibfnamefont {H.}~\bibnamefont {Ji}},
  \bibinfo {author} {\bibfnamefont {N.}~\bibnamefont {Yao}}, \bibinfo {author}
  {\bibfnamefont {J.~J.}\ \bibnamefont {Clarke}}, \bibinfo {author}
  {\bibfnamefont {R.~J.}\ \bibnamefont {Cava}}, \ and\ \bibinfo {author}
  {\bibfnamefont {J.~R.}\ \bibnamefont {Petta}},\ }\href {\doibase
  10.1063/1.4892353} {\bibfield  {journal} {\bibinfo  {journal} {Appl. Phys.
  Lett.}\ }\textbf {\bibinfo {volume} {105}},\ \bibinfo {pages} {053512}
  (\bibinfo {year} {2014})}\BibitemShut {NoStop}%
\bibitem [{\citenamefont {Lang}\ \emph {et~al.}(2014)\citenamefont {Lang},
  \citenamefont {Montazeri}, \citenamefont {Onbasli}, \citenamefont {Kou},
  \citenamefont {Fan}, \citenamefont {Upadhyaya}, \citenamefont {Yao},
  \citenamefont {Liu}, \citenamefont {Jiang}, \citenamefont {Jiang},
  \citenamefont {Wong}, \citenamefont {Yu}, \citenamefont {Tang}, \citenamefont
  {Nie}, \citenamefont {He}, \citenamefont {Schwartz}, \citenamefont {Wang},
  \citenamefont {Ross},\ and\ \citenamefont {Wang}}]{Wang2014}%
  \BibitemOpen
  \bibfield  {author} {\bibinfo {author} {\bibfnamefont {M.}~\bibnamefont
  {Lang}}, \bibinfo {author} {\bibfnamefont {M.}~\bibnamefont {Montazeri}},
  \bibinfo {author} {\bibfnamefont {M.~C.}\ \bibnamefont {Onbasli}}, \bibinfo
  {author} {\bibfnamefont {X.}~\bibnamefont {Kou}}, \bibinfo {author}
  {\bibfnamefont {Y.}~\bibnamefont {Fan}}, \bibinfo {author} {\bibfnamefont
  {P.}~\bibnamefont {Upadhyaya}}, \bibinfo {author} {\bibfnamefont
  {K.}~\bibnamefont {Yao}}, \bibinfo {author} {\bibfnamefont {F.}~\bibnamefont
  {Liu}}, \bibinfo {author} {\bibfnamefont {Y.}~\bibnamefont {Jiang}}, \bibinfo
  {author} {\bibfnamefont {W.}~\bibnamefont {Jiang}}, \bibinfo {author}
  {\bibfnamefont {K.~L.}\ \bibnamefont {Wong}}, \bibinfo {author}
  {\bibfnamefont {G.}~\bibnamefont {Yu}}, \bibinfo {author} {\bibfnamefont
  {J.}~\bibnamefont {Tang}}, \bibinfo {author} {\bibfnamefont {T.}~\bibnamefont
  {Nie}}, \bibinfo {author} {\bibfnamefont {L.}~\bibnamefont {He}}, \bibinfo
  {author} {\bibfnamefont {R.~N.}\ \bibnamefont {Schwartz}}, \bibinfo {author}
  {\bibfnamefont {Y.}~\bibnamefont {Wang}}, \bibinfo {author} {\bibfnamefont
  {C.~A.}\ \bibnamefont {Ross}}, \ and\ \bibinfo {author} {\bibfnamefont
  {K.~L.}\ \bibnamefont {Wang}},\ }\href {\doibase 10.1021/nl500973k}
  {\bibfield  {journal} {\bibinfo  {journal} {Nano Lett.}\ }\textbf {\bibinfo
  {volume} {14}},\ \bibinfo {pages} {3459} (\bibinfo {year}
  {2014})}\BibitemShut {NoStop}%
\bibitem [{\citenamefont {Huang}\ \emph {et~al.}(2017)\citenamefont {Huang},
  \citenamefont {Chong}, \citenamefont {Tung}, \citenamefont {Chen},
  \citenamefont {Wu}, \citenamefont {Lee}, \citenamefont {Huang}, \citenamefont
  {Li},\ and\ \citenamefont {Qiu}}]{Qiu2017}%
  \BibitemOpen
  \bibfield  {author} {\bibinfo {author} {\bibfnamefont {S.-Y.}\ \bibnamefont
  {Huang}}, \bibinfo {author} {\bibfnamefont {C.-W.}\ \bibnamefont {Chong}},
  \bibinfo {author} {\bibfnamefont {Y.}~\bibnamefont {Tung}}, \bibinfo {author}
  {\bibfnamefont {T.-C.}\ \bibnamefont {Chen}}, \bibinfo {author}
  {\bibfnamefont {K.-C.}\ \bibnamefont {Wu}}, \bibinfo {author} {\bibfnamefont
  {M.-K.}\ \bibnamefont {Lee}}, \bibinfo {author} {\bibfnamefont {J.-C.-A.}\
  \bibnamefont {Huang}}, \bibinfo {author} {\bibfnamefont {Z.}~\bibnamefont
  {Li}}, \ and\ \bibinfo {author} {\bibfnamefont {H.}~\bibnamefont {Qiu}},\
  }\href {\doibase 10.1038/s41598-017-02662-8} {\bibfield  {journal} {\bibinfo
  {journal} {Sci. Rep.}\ }\textbf {\bibinfo {volume} {7}},\ \bibinfo {pages}
  {2422} (\bibinfo {year} {2017})}\BibitemShut {NoStop}%
\end{thebibliography}%

\end{document}